\documentclass[a4paper]{pasj01}

\usepackage{color}
\usepackage{graphicx}

\begin{document} 


\Received{2018/07/17}
\Accepted{2018/09/30}
\Published{2018/mm/dd}

\title{Detection of polarized gamma-ray emission from the Crab nebula with Hitomi Soft Gamma-ray Detector
\thanks{Corresponding authors are
Shin \textsc{Watanabe},
Yuusuke \textsc{Uchida},
Hirokazu \textsc{Odaka},
Greg \textsc{Madejski},
Katsuhiro \textsc{Hayashi},
Tsunefumi \textsc{Mizuno},
Rie \textsc{Sato},
and
Yoichi \textsc{Yatsu}.
}
}

\author{Hitomi Collaboration,
Felix \textsc{Aharonian}\altaffilmark{1,2,3},
Hiroki \textsc{Akamatsu}\altaffilmark{4},
Fumie \textsc{Akimoto}\altaffilmark{5},
Steven W. \textsc{Allen}\altaffilmark{6,7,8},
Lorella \textsc{Angelini}\altaffilmark{9},
Marc \textsc{Audard}\altaffilmark{10},
Hisamitsu \textsc{Awaki}\altaffilmark{11},
Magnus \textsc{Axelsson}\altaffilmark{12},
Aya \textsc{Bamba}\altaffilmark{13,14},
Marshall W. \textsc{Bautz}\altaffilmark{15}
% ...
}

\altaffiltext{1}{Dublin Institute for Advanced Studies, 31 Fitzwilliam Place, Dublin 2, Ireland}
\altaffiltext{2}{Max-Planck-Institut f{\"u}r Kernphysik, P.O. Box 103980, 69029 Heidelberg, Germany}
\altaffiltext{3}{Gran Sasso Science Institute, viale Francesco Crispi, 7 67100 L'Aquila (AQ), Italy}
\altaffiltext{4}{SRON Netherlands Institute for Space Research, Sorbonnelaan 2, 3584 CA Utrecht, The Netherlands}
\altaffiltext{5}{Institute for Space-Earth Environmental Research, Nagoya University, Furo-cho, Chikusa-ku, Nagoya, Aichi 464-8601}
\altaffiltext{6}{Kavli Institute for Particle Astrophysics and Cosmology, Stanford University, 452 Lomita Mall, Stanford, CA 94305, USA}
\altaffiltext{7}{Department of Physics, Stanford University, 382 Via Pueblo Mall, Stanford, CA 94305, USA}
\altaffiltext{8}{SLAC National Accelerator Laboratory, 2575 Sand Hill Road, Menlo Park, CA 94025, USA}
\altaffiltext{9}{NASA, Goddard Space Flight Center, 8800 Greenbelt Road, Greenbelt, MD 20771, USA}
\altaffiltext{10}{Department of Astronomy, University of Geneva, ch. d'\'Ecogia 16, CH-1290 Versoix, Switzerland}
\altaffiltext{11}{Department of Physics, Ehime University, Bunkyo-cho, Matsuyama, Ehime 790-8577}
\altaffiltext{12}{Department of Physics and Oskar Klein Center, Stockholm University, 106 91 Stockholm, Sweden}
\altaffiltext{13}{Department of Physics, The University of Tokyo, 7-3-1 Hongo, Bunkyo-ku, Tokyo 113-0033}
\altaffiltext{14}{Research Center for the Early Universe, School of Science, The University of Tokyo, 7-3-1 Hongo, Bunkyo-ku, Tokyo 113-0033}
\altaffiltext{15}{Kavli Institute for Astrophysics and Space Research, Massachusetts Institute of Technology, 77 Massachusetts Avenue, Cambridge, MA 02139, USA}
% ...

\email{watanabe@astro.isas.jaxa.jp}


\maketitle

\begin{abstract}
We present the results from the Hitomi Soft Gamma-ray Detector (SGD) observation of the Crab nebula. The main part of SGD is a Compton camera, 
which in addition to being a spectrometer, is capable of measuring polarization of gamma-ray photons. The Crab nebula is 
one of the brightest X-ray / gamma-ray sources on the sky, and, the only source from which polarized X-ray photons have been detected.
SGD observed the Crab nebula during the initial test observation phase of Hitomi. 
We performed the data analysis of the SGD observation, the SGD background estimation and the SGD Monte Carlo simulations, and, successfully 
detected polarized gamma-ray emission from the Crab nebula with only about 5~ks exposure time. The obtained 
polarization fraction of the phase-integrated Crab emission (sum of pulsar and nebula emissions) 
is (22.1 $\pm$ 10.6)\% and, the polarization angle is 110.7\degree + 13.2\degree / $-$13.0\degree~ 
in the energy range of 60--160~keV (The errors correspond to the 1 sigma deviation). 
The confidence level of the polarization detection was 99.3\%. 
The polarization angle measured by SGD is about one sigma deviation with the projected spin axis of the pulsar, 124.0\degree$\pm$0.1\degree.
\end{abstract}

\newpage

\section{Introduction}

In addition to spectral, temporal, and imaging information gleaned from observations 
of any astrophysical sources, polarization of electromagnetic emission from those sources 
provides the fourth handle on understanding the operating radiative processes.  
Historically, measurement of high radio polarization from celestial sources implicated 
synchrotron radiation as such process, first suggested by \citet{Shklovsky:1970}.  Measurement of 
radio or optical polarization is relatively straightforward:  first, it can be done from the 
Earth's surface, and second, the instruments are relatively simple.
The measurements in the X-ray band are more complicated:  those have to be conducted 
from space which constrains the instrument size, and, unlike e.g. radio waves, 
X-rays are usually detected as particles and require large statistics to measure the polarization.

One of the brightest X-ray sources on the sky, with appreciable polarization measured 
in the radio and optical bands is the Crab nebula.  It was detected by (probably) every orbiting 
X-ray astronomy mission (for a recent summary, see \cite{Hester:2008}).  It was thus expected that 
X-ray polarization should be detected as well, and in fact, the first instrument sensitive to 
X-ray polarization, the OSO-8 mission, observed the Crab nebula, and detected X-ray polarization 
\citep{Weisskopf:1978}.  The measurement, performed at 2.6~keV, measured polarization at 
roughly $\sim 20 \pm 1$\% level.  It was some 30 years later that the {\sl INTEGRAL} 
mission observed the Crab nebula and detected significant polarization of its hard X-ray / soft 
$\gamma$-ray emission \citep{Chauvin:2013, Forot:2008}. 
Moreover, {\sl INTEGRAL} teams reported gamma-ray polarization measurements 
from the black hole binary system Cygnus X-1 \citep{Laurent:2010, Jourdain:2012, Rodriguez:2015}.
However, the interpretation 
of the measurements with {\sl INTEGRAL} are not straightforward, because its instruments 
were not designed for, or calibrated for polarization measurements.

More recently, the Crab nebula was observed by the balloon-borne mission PoGOLite Pathfinder 
\citep{Chauvin:2016}, and PoGO+\citep{Chauvin:2017, Chauvin:2018}, 
with clear detection of soft $\gamma$-ray polarization in the $\sim 18 - 160$~keV band, 
thus expanding the X-ray band where the Crab nebula emission shows polarization.  
The PoGO+ is an instrument employing a plastic scintillator, with an effective area of 
378 cm$^{2}$ and optimized for polarization measurements of Compton scattering perpendicular to the incident direction where the modulation factor of the azimuth scattering angle is high;
the PoGO+ team reports the polarization of the phase-integrated Crab emission of
$20.9 \pm 5.0$\% with a polarization angle of 131.3\degree~$\pm$~6.8\degree, 
while in the off-pulse phase, it is $17.4^{+8.6}_{-9.3}$ \% with a polarization angle of 137\degree~$\pm$~15\degree~.

The Japanese mission Hitomi \citep{Takahashi:2018}, launched in 2016, included the 
Soft Gamma-ray Detector (SGD), an instrument sensitive in the 60--600~keV range, 
but also capable of measuring polarization (see \cite{Tajima:2018}) since it employs a Compton camera as a gamma-ray detector.
The SGD was primarily designed as a spectrometer, but, it was also optimized for polarization measurements
(see, e.g., \cite{Tajima:2010}). 
For example, the Compton camera of the SGD is highly efficient for Compton scattering perpendicular to the incident photon direction and is symmetric with 90\degree~ rotation.
The calibration and the performance verification as a polarimeter have already been performed 
by using polarized soft gamma-ray beam at SPring-8 \citep{Katsuta:2016}.
Hitomi did observe the Crab nebula in the early phase of the mission. 
Since the goal of the observation reported here was to verify the performance of Hitomi's instruments rather than to perform detailed scientific studies of the Crab nebula, 
the observation time was short.
Even though this observation was conducted during orbits where the satellite passed through the high-background orbital regions including orbits crossing the South Atlantic Anomaly, 
the Crab nebula was still readily detected, as we report in subsequent sections.
We discuss the data reduction and analysis in section~2 and section~3, 
the measurement of Crab's polarization in section~4, compare 
our measurement to previous measurements in section~5, and also discuss the implications 
on the modeling of the Crab nebula in section~5.  We note that the Crab nebula observations with Hitomi's 
Soft X-ray Spectrometer were published recently \citep{Hitomi:2018SXS}, and 
observations with the Hard X-ray Imager are in preparation. Moreover, the data analysis of the Crab pulsar with Hitomi's instruments
were also published \citep{Hitomi:2018Crabpulsar}.

\section{Crab Observation with SGD}

\subsection{Instrument and Data Selection}
The Soft Gamma-ray Detector (SGD) was one of the instruments 
deployed on the Hitomi satellite (see \cite{Takahashi:2018} for the detailed description 
of the Hitomi mission).  The instrument was a collimated Si/CdTe 
Compton camera with the field of view of 0.6\degree~$\times$~0.6\degree, 
sensitive in the 60--600~keV band; for details of the SGD, see \citet{Tajima:2018}.
The SGD Compton camera consisted of 32 layers of Si pixel sensors where Compton scatterings take place primarily.
Each layer of the Si sensor had a $16\times16$ array of $3.2\times3.2$~mm$^2$ pixels with a thickness of 0.6~mm.
In order to efficiently detect photons scattered in the Si sensor stack, it was surrounded on 5 sides by 0.75~mm thick CdTe pixel sensors where photo-absorptions take place primarily.
In the forward direction, 8 layers of CdTe sensors with a $16\times16$ array of $3.2\times3.2$~mm$^2$ pixels were placed, 
while 2 layers of CdTe sensors with $16\times24$ array of $3.2\times3.2$~mm$^2$ pixels were placed on four sides of the Si sensor stack.
For details of SGD Compton camera, see \citet{Watanabe:2014}.
The SGD consisted 
of two detector units, SGD1 and SGD2, each containing three Compton cameras, named 
as CC1, CC2, and CC3, respectively. Those detectors were surrounded on five sides 
by an anti-coincidence detector containing BGO scintillator. The observation of the 
Crab nebula with Hitomi was performed from 12:35 to 18:01 UT on March 25, 2016.  
This observation followed the start-up operations for the SGD, which were held from 
March 15 to March 24, and, all cameras of both SGD1 and SGD2 went into the nominal 
observation mode before the Crab nebula observation.  However, just before the Crab nebula
observation it was found that one channel in the CdTe detectors of SGD2 
CC2 became noisy, and subsequently we set the voltage value of the high-voltage power supply for the CdTe sensors of SGD2 CC2 to 0~V during the Crab nebula observation.
Since CC3 shares the same high-voltage power supply with CC2, the CdTe sensors in CC3 are also disabled.
Therefore, four of six Compton cameras (SGD1 CC1, CC2, CC3 and SGD2 CC1) were operated in the 
nominal mode, which enabled the Compton event reconstruction. 

Good time intervals (GTI) of SGD during the Crab observation are listed 
in Table~\ref{tab:crabgti}.  The intervals during the Earth occultation 
and South Atlantic Anomaly (SAA) passages are excluded.
The total on-source duration was 8.6~ks.   The exposure times of each Compton 
camera after dead-time corrections are listed in Table~\ref{tab:crabexposure}.
In the SGD1 Compton cameras, the dead-time corrected exposure time can be 
derived from the number of ``clean" pseudo events 
\citep{Watanabe:2014}, which have no {\tt FBGO} flag and no {\tt HITPATBGO} flag. 
The pseudo events are events triggered by ``pseudo triggers", which are generated randomly 
in the Compton camera FPGA based on the pseudorandom numbers calculated in the FPGA.
The count rate of the pseudo triggers is set to be 2~Hz.
{\tt FBGO} and {\tt HITPATBGO} flags indicate existence of anti-coincidence signals from the BGO shield.
The pseudo events are processed in the same manner as usual triggers, and, are discarded 
if the pseudo trigger is generated while a ``real event" is inhibiting other triggers. 
Therefore, the dead-time fraction can be estimated by counting a number of pseudo events,
and, the dead-time by accidental hits in BGOs can be also estimated from the pseudo events 
with {\tt FBGO} flags and {\tt HITPATBGO} flags.
However, it was found that there was an error in the on-board readout logic of adding 
the {\tt HITPAT} BGO flags to pseudo events for the parameter setting 
of SGD2 CC1.  Due to this error, dead-time fraction by the accidental hit in the 
BGOs cannot be derived from the number of pseudo events generated from SGD2 CC1.
Therefore, for SGD2 CC1, the dead-time fraction due to accidental hit in BGOs was 
calculated from the fraction of ``clean" pseudo events in the SGD2 CC2, 
allowing the determination of the dead-time corrected exposure time.  
For SGD2 CC2, a parameter setting to avoid the error has been used. And, 
the dead-time fraction by accidental hits in BGOs must be same among the Compton cameras in SGD2, 
because the BGO signals are common among all three Compton cameras in SGD2.

The attitude of the Hitomi satellite was stable throughout the Crab GTI. The nominal 
pointing position is (R.A., DEC.) = (83.6334\degree, 22.0132\degree) and the 
nominal roll angle is 267.72\degree\, that is measured from the north to the satellite Y axis counter-clockwise. 
The distance from the nominal pointing 
position is within 0.3~arcmin for the 98.7\% of the observation time. 
The difference from the nominal roll angle is within 0.05\degree~ for the 
99.6\% of the observation time.  Therefore, these offsets from the true direction of 
Crab are negligible and we have not considered them in the analysis.
 
\begin{table*}
  \tbl{The good time intervals of the Crab observation. }{%
  \begin{tabular}{ccccc}
      \hline
      TSTART [s]\footnotemark[$\dag$] & TSTART [UTC] & TSTOP [s]\footnotemark[$\dag$] & TSTOP [UTC] & duration [s]\\ 
      \hline
	   70374949.000000 & 2016/03/25 12:35:48 & 70374979.000000 & 2016/03/25 12:36:18 & 30\\
	   70375027.000000 & 2016/03/25 12:37:06 & 70377352.000000 & 2016/03/25 13:15:51 & 2325\\
	   70380742.000000 & 2016/03/25 14:12:21 & 70383114.000000 & 2016/03/25 14:51:53 & 2372\\
	   70386733.000000 & 2016/03/25 15:52:12 & 70388875.000000 & 2016/03/25 16:27:54 & 2142\\
	   70392719.000000 & 2016/03/25 17:31:58 & 70394479.234375 & 2016/03/25 18:01:18.234375 & 1760\\
      \hline
    \end{tabular}}\label{tab:crabgti}
\begin{tabnote}
\footnotemark[$\dag$] : TSTART and TSTOP is expressed in AHTIME, defined as the time elapsed since 2014/01/01 00:00:00 in seconds.
\end{tabnote}
\end{table*}

\begin{table*}
  \tbl{Exposures of the Crab observation.}{%
  \begin{tabular}{cccccc}
      \hline
       & No. of all pseudo & No. of ``clean" pseudo & Live Time          & dead time fraction            & Live Time  \\
       &                   &                       &  from clean pseudo & due to BGO accidental hits &  for SGD2 CC1     \\
      \hline
      SGD1 CC1 & 11084 & 9879 & 4939.5~s & & \\
      SGD1 CC2 & 10624 & 9478 & 4739.0~s & & \\
      SGD1 CC3 & 11036 & 9879 & 4939.5~s & & \\
      SGD2 CC1 & 11826 &      &          & 0.1161 & 5226.29~s \\
      SGD2 CC2 & 11788 & 10419 & 5209.5~s & 0.1161\footnotemark[$\dag$] & \\ 
      \hline
    \end{tabular}}\label{tab:crabexposure}
\begin{tabnote}
\footnotemark[$\dag$] : this value is derived from the number of all pseudo events and the number of ``clean" pseudo events. $\left[ (11788-10419)/11788 \right]$
\end{tabnote}
\end{table*}

\subsection{Background Determination}

\begin{figure}[h]
 \begin{center}
  \includegraphics[width=0.45\textwidth]{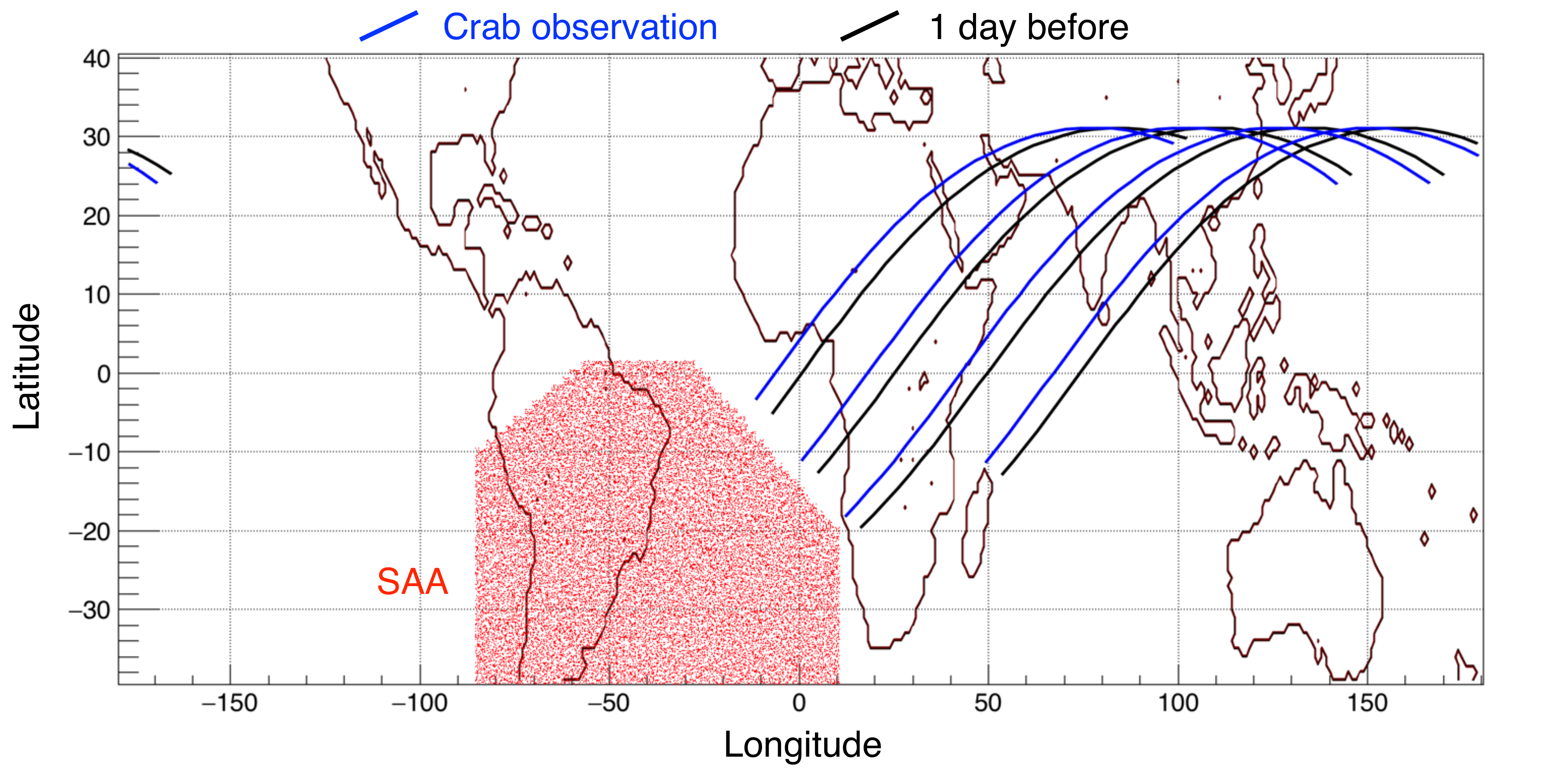} 
 \end{center}
\caption{The satellite position during observations. The black line shows 
the satellite position during the Crab GTI, and the blue line shows the position 
during the epoch one day earlier Crab GTI.}\label{fig:satpos_crabobs}
\end{figure}

Figure~\ref{fig:satpos_crabobs} shows the Hitomi satellite position during the Crab GTI and one day before the Crab GTI,
when the satellite was pointing at RXJ~1856.5$-$3754, which is a very weak source in the hard X-ray/soft gamma-ray band, 
and thus such "one day earlier" observation is a good proxy to measure the background. 
The time interval information about observations performed one day earlier than the Crab GTI are listed in Table~\ref{tab:crabgti_onedayearlier}.
Because the observations start soon after the SAA passages, 
the background rate during the Crab GTI was higher than the average due to short-lived activated materials produced in the SAA.
Although the Crab nebula is one of the brightest sources in this energy region, the background events were not negligible for spectral analysis and polarization measurements.
As shown in Figure~\ref{fig:satpos_crabobs}, the satellite positions and the orbit conditions one day earlier than the Crab GTI are 
similar to those during the Crab GTI, which would imply background conditions could be similar.

\begin{table*}
  \tbl{The time intervals of pointings performed one day earlier than the Crab GTI.}{%
  \begin{tabular}{ccccc}
      \hline
      TSTART [s]\footnotemark[$\dag$] & TSTART [UTC] & TSTOP [s]\footnotemark[$\dag$] & TSTOP [UTC] & duration [s]\\ 
      \hline
	   70288549.000000 & 2016/03/24 12:35:48 & 70288579.000000 & 2016/03/24 12:36:18 & 30\\
	   70288627.000000 & 2016/03/24 12:37:06 & 70290952.000000 & 2016/03/24 13:15:51 & 2325\\
	   70294342.000000 & 2016/03/24 14:12:21 & 70296714.000000 & 2016/03/24 14:51:53 & 2372\\
	   70300333.000000 & 2016/03/24 15:52:12 & 70302475.000000 & 2016/03/24 16:27:54 & 2142\\
	   70306319.000000 & 2016/03/24 17:31:58 & 70308079.234375 & 2016/03/24 18:01:18.234375 & 1760\\
      \hline
    \end{tabular}}\label{tab:crabgti_onedayearlier}
\begin{tabnote}
\footnotemark[$\dag$] : TSTART and TSTOP is expressed in AHTIME, defined as the time elapsed since 2014/01/01 00:00:00 in seconds.
\end{tabnote}
\end{table*}

\begin{figure}[h]
 \begin{center}
  \includegraphics[width=0.4\textwidth]{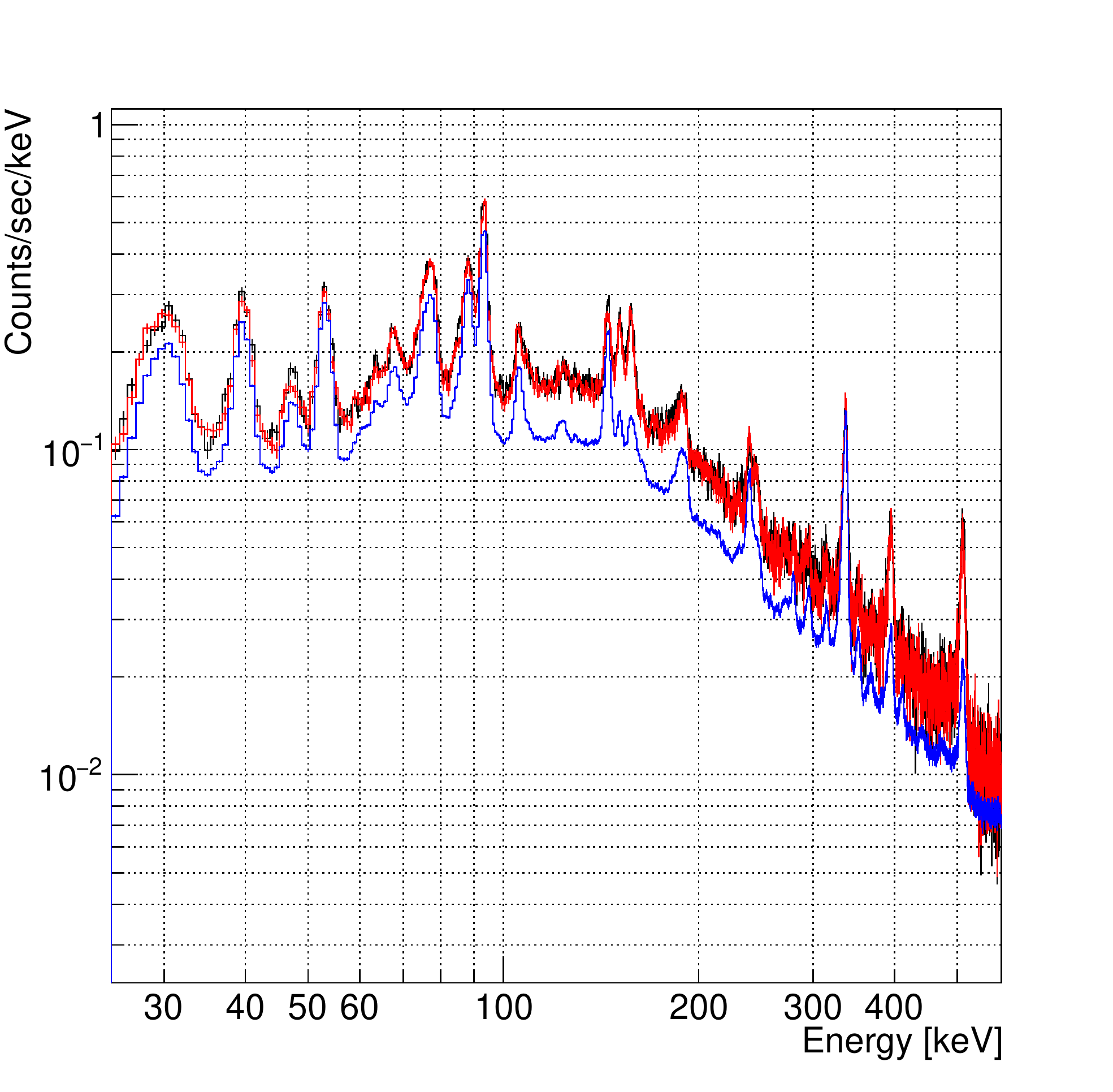} 
 \end{center}
\caption{Spectra of CdTe side single hit events. The red and the black show the spectra for the one day 
and two days earlier than the Crab GTI, respectively. 
The blue spectrum shows the single hit events of CdTe-side sensors on the 
orbit that the satellite does not pass the SAA region. }\label{fig:cdtes_bgd}
\end{figure}

In order to confirm that the satellite encountered the similar background environments during similar orbit conditions, 
we compare the SGD data between an epoch one day earlier and also two days earlier than the Crab observation GTIs.
The single hit spectra obtained by the CdTe-side sensors are shown in Figure~\ref{fig:cdtes_bgd}.
The CdTe-side sensors are located on the four sides around the stack of Si/CdTe sensors inside the Compton camera,
and, are not exposed to gamma rays from the field of view.
Therefore, the influence of the background environment should be reflected strongly in the single hit event in the CdTe-side detectors.
The red and the black points show the spectra for the epochs one day and two days 
earlier than the Crab GTI,  
respectively. These two spectra have the same spectral shape including various emission 
lines from activated materials. The flux levels were the same within 3\%. On the other hand, 
the blue spectrum shows the single hit events of CdTe-side detectors on the orbit where the 
satellite does not pass the SAA region.  Although the background environment 
varied during one day, it was found that the background estimation becomes possible by 
using the data from one day earlier.

\begin{figure}[h]
 \begin{center}
  \includegraphics[width=0.45\textwidth]{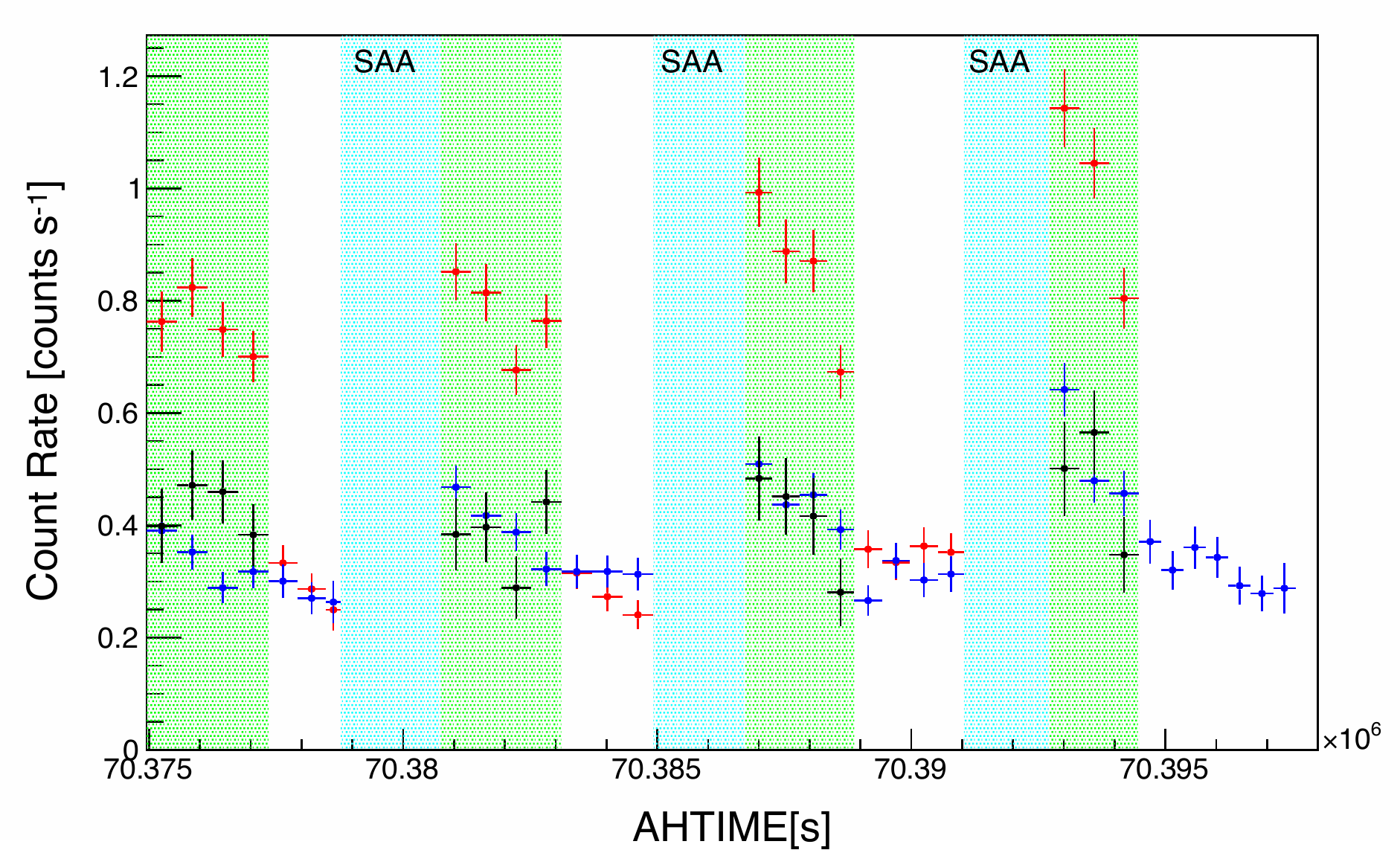} 
 \end{center}
\caption{Count rate the SGD Compton camera as a function of time. 
The red and the blue points show the count rates during the Crab observation and one day earlier.
The black points show the count rates of the Crab GTI after subtracting the count rates one day earlier.
The regions filled in green show the Crab GTI. The regions filled in cyan show time intervals excluded from the GTI due to the SAA passages.
In the  "white" portions of time intervals, the Crab nebula was not able to be observed because of the Earth occultation.
}\label{fig:lightcurve_crab}
\end{figure}

In order to further verify the background subtraction using the data one day earlier, the count rates as a function of the time 
during the Crab GTI and one day earlier are compared in Figure~\ref{fig:lightcurve_crab}.
The red and the blue points show the count rates during the Crab GTI and one day earlier.
The black points show the count rates of the Crab GTI after subtracting the count rates one day earlier, which corresponds to the count rates of the Crab nebula.
Since the black points do not show any visible systematic trend implying additional backgrounds, 
it implies this background subtraction is appropriate.

\newpage

\section{Data Analysis}

\subsection{Data Processing with Hitomi tools}
The data processing and the event reconstruction are performed by the standard 
Hitomi pipeline using the Hitomi 
{\tt ftools}\citep{Angelini:2018}\footnote{https://heasarc.gsfc.nasa.gov/lheasoft/ftools/headas/hitomi.html}.
In the pipeline process for SGD, the ftools used for the SGD are \texttt{hxisgdsff}, 
\texttt{hxisgdpha} and \texttt{sgdevtid}.  
The \texttt{hxisgdsff} converts the raw event data into the predefined data format.
The \texttt{hxisgdpha} calibrates the event energy. 
The \texttt{sgdevtid} reconstruct each event.
These tools are included in HEASoft after version is 6.19.
The version of the calibration files used in these process is 20140101v003.

The \texttt{sgdevtid} is one of key tools for the SGD event reconstruction, 
which determines whether the sequence of interactions is valid 
and computes the event energy and the 3-dimensional coordinates of its first interaction.
The event reconstruction procedure of the \texttt{sgdevtid} is described in \cite{Ichinohe:2016}.
The first step of the process is to merge signals that are consistent with fluorescence X-rays with the original interaction sites according to their locations and energies. 
The merging process combines the separated signals into a hit for each interaction.
The second step is to analyze the reconstructed hits and determine whether the sequence is consistent with an event.
This step depends on the number of reconstructed hits.
If there is only one hit, the process is done, and, the energy information and the hit position information are recorded on the output event file 
as a ``single hit" event. 
In the case of the event which has 2 to 4 hits, the process determines whether the event is a valid gamma-ray event and whether the first interaction is Compton scattering 
by applying the Compton kinematics equation:
\begin{equation}
\cos \theta_\mathrm{K} = 1 - m_\mathrm{e}c^2\left(\frac{1}{(E_\gamma-E_1)} - \frac{1}{E_\gamma}\right),
\end{equation}
where $\theta_\mathrm{K}$ is scattering angle defined by Compton kinematics, $m_\mathrm{e}c^2$ is the rest energy of an electron, 
$E_1$ is the first hit energy corresponding to the recoil energy of the scattered electron 
and $E_\gamma$ is the reconstructed energy of the incoming gamma-ray photon.
All possible permutations for the sequence of hits are tried and 
all sequences with non-physical Compton scattering angle ($|\cos \theta_\mathrm{K}| > 1$) are rejected.
Besides the kinematic scattering angle $\theta_\mathrm{K}$, the geometrical scattering angles $\theta_{geometry}$ can be derived from 
the directions of the incident gamma ray and the scattered gamma ray. 
The incident gamma ray is assumed to be aligned with the line of sight. 
The direction of the scattered gamma ray is reconstructed from the positions of the first and the second hits.
The difference of them is called angular resolution measure (ARM):
\begin{equation}
\mathrm{ARM}\  := \theta_\mathrm{K} - \theta_{geometry}.
\end{equation}
If more than one sequence remains, the order of hits with the smallest ARM value is selected as the most likely sequence.
Moreover, in the case of 3-hit events, the second interaction is assumed to be the Compton scattering, and, in the case of 4-hit events, 
the second and the third interactions are assumed to be the Compton scatterings.
For these interactions, the tests of Compton kinematics and differences between kinematic scattering angles and geometrical scattering angles 
are performed. If the sequences have any non-physical Compton scatterings or any inconsistent kinematic angles to the geometric scattering angles,
the sequences are rejected.
In the first calculation, the reconstructed energy of the incoming gamma-ray photon $E_\gamma$ is set to be
\begin{equation}
E_\gamma = \sum_i E_i,
\end{equation}
where $E_i$ is the energy information of the i-th hit. For 3-hit and 4-hit events, if all sequences are rejected in this calculation, 
the \texttt{sgdevtid} calculates the escape energy, the unabsorbed part of the energy of a photon that is able to exit the camera after detections, 
and executes the previous tests again. 
Finally, for such good ``Compton event" after the process, the information for the first interaction such 
as $\cos\theta_\mathrm{K}$, the azimuthal angle $\phi$ of scattered gamma-rays, and the ARM value as `\texttt{OFFAXIS}'
are recorded on the output event file in addition to the reconstructed energy information and the first hit position information
\footnote{The details of recorded columns are shown in https://heasarc.gsfc.nasa.gov/ftools/caldb/help/sgdevtid.html}.

\subsection{Processing of Crab Observation Data}

\begin{figure*}[h]
 \begin{center}
  \includegraphics[width=6cm]{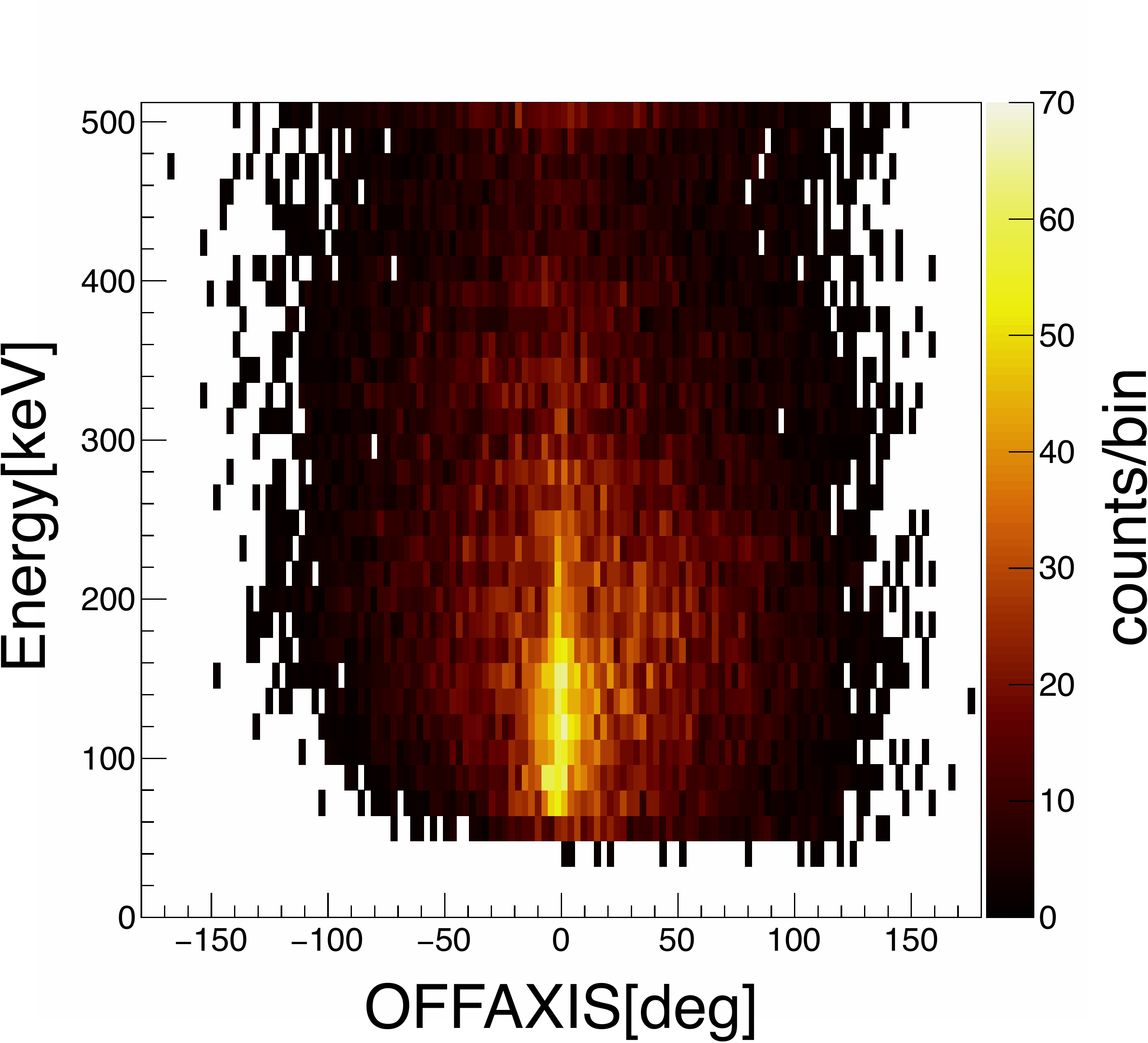} 
  \includegraphics[width=6cm]{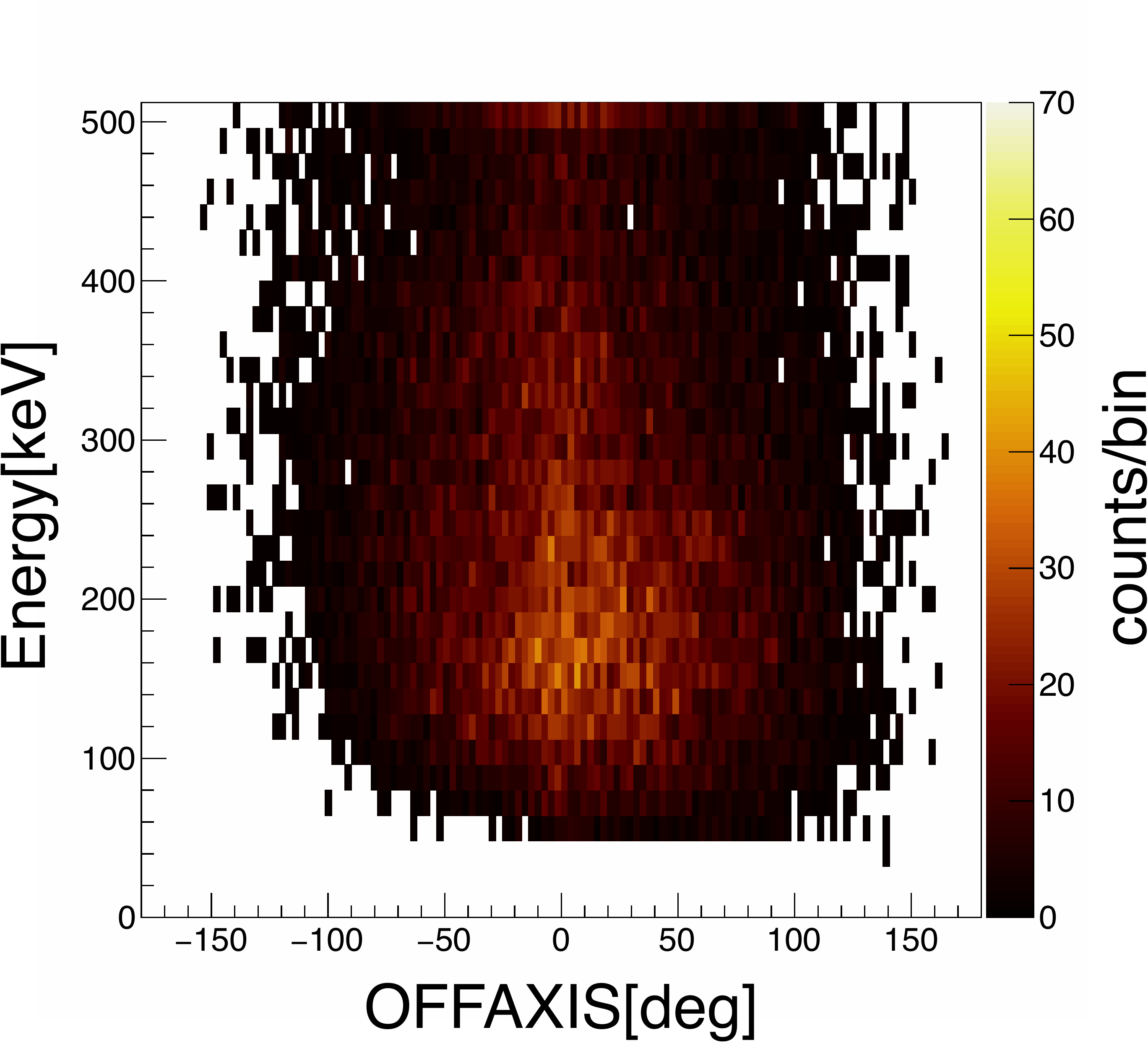} 
 \end{center}
\caption{Two-dimensional histograms of Compton-reconstructed events.  
The relation between \texttt{OFFAXIS} and energy is shown. The left-hand panel is 
the histogram made from the events during the Crab GTI, and, the right-hand 
panel is prepared from the events collected one day earlier than the Crab GTI.}\label{fig:h2dim_crab}
\end{figure*}

Figure~\ref{fig:h2dim_crab} shows a relation between \texttt{OFFAXIS} and energy spectrum for 
the ``Compton-reconstructed" events where \texttt{sgdevtid} finds the position of 
the first Compton scattering with physical $\cos\theta_\mathrm{K}$ in Si sensors.
The histogram in the left-hand panel is made from the events during the Crab GTI, 
and, that in the right-hand panel is made from the events collected one day earlier 
than the Crab GTI.  An excess at around \texttt{OFFAXIS} $\sim$~0\degree~ can be seen in the 
histogram of the Crab GTI corresponding to the gamma rays from the Crab nebula.

\begin{figure*}[h]
 \begin{center}
  \includegraphics[width=5cm]{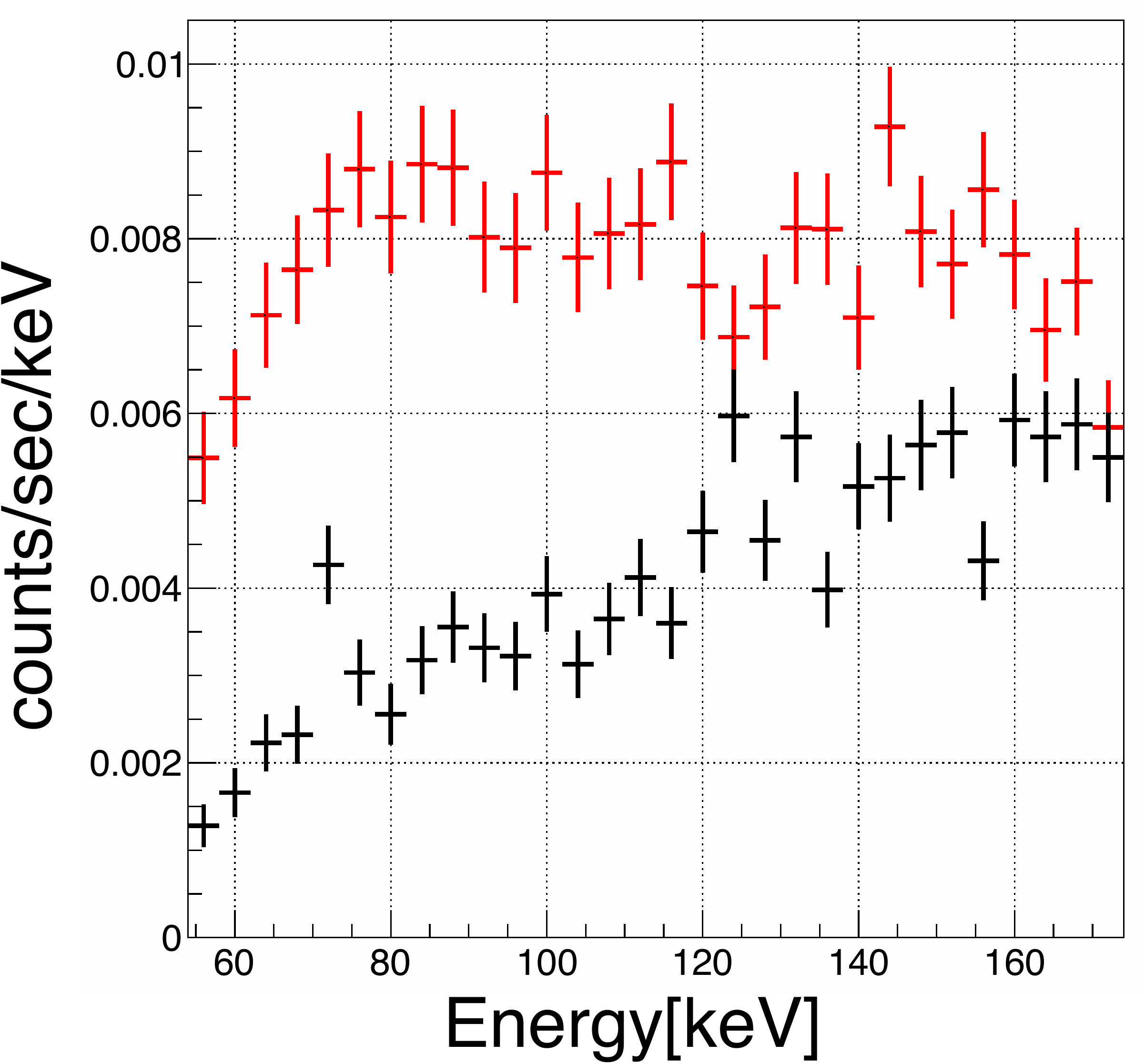} 
  \includegraphics[width=5cm]{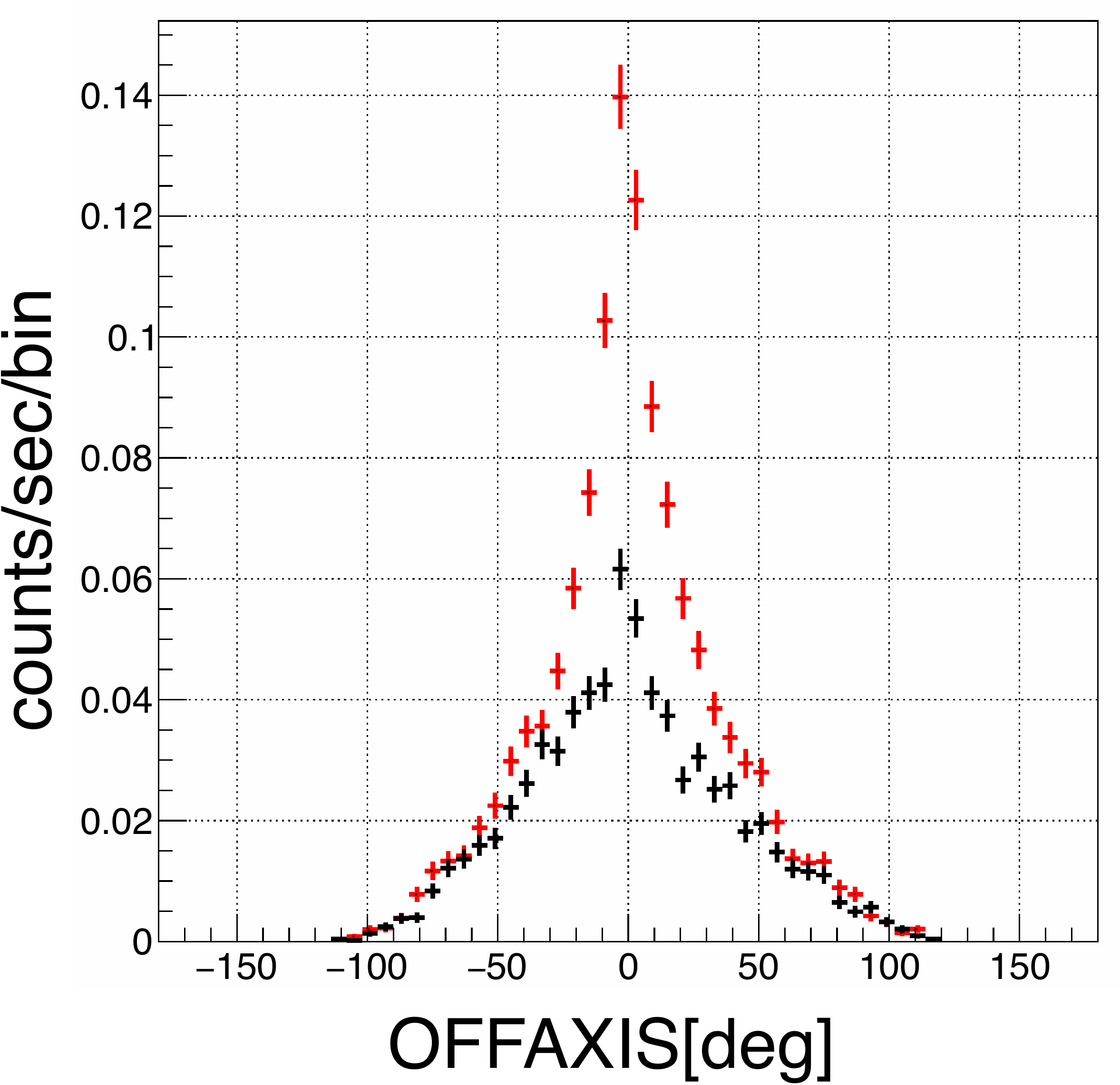} 
  \includegraphics[width=5cm]{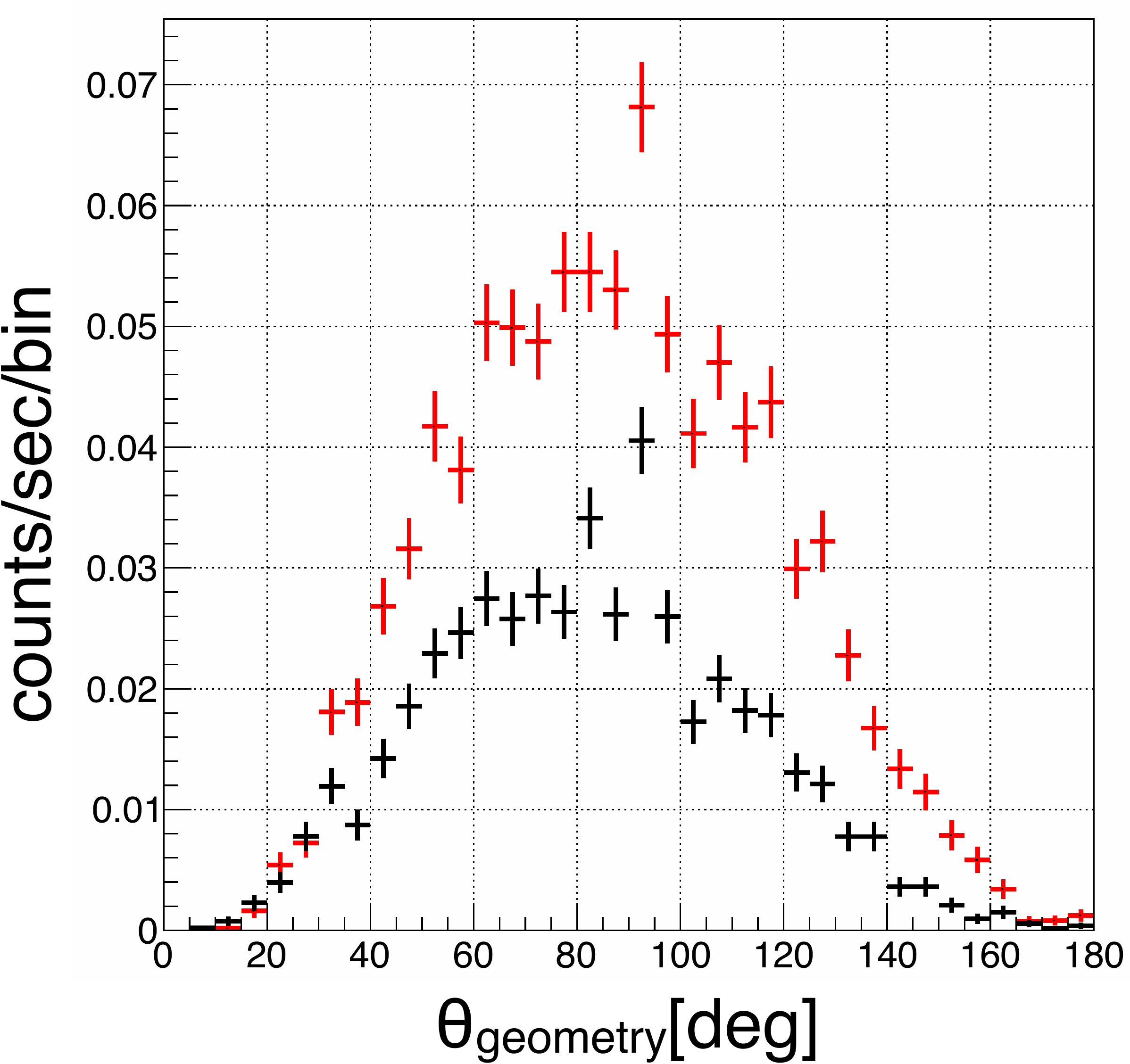} 
 \end{center}
\caption{The histograms of Energy, \texttt{OFFAXIS}, $\theta_{geometry}$. The selection criteria 
are 60~keV~$<$~Energy~$<$~160~keV, $-$30\degree~$<$~\texttt{OFFAXIS}~$<$~$+$30\degree~ and 
50\degree~$<$~$\theta_{geometry}$~$<$~150\degree. The red histograms are made from 
the events during the Crab GTI, and the black ones are from the events during the 
epoch one day earlier than the Crab GTI.}\label{fig:crab_obs}
\end{figure*}

In order to obtain good signal to noise ratio, selections of 
60~keV~$<$~Energy~$<$~160~keV, $-$30\degree~$<$~\texttt{OFFAXIS}~$<$~$+$30\degree, 
50\degree~$<$~$\theta_{geometry}$~$<$~150\degree~ are applied.
The histograms of Energy, \texttt{OFFAXIS}, $\theta_{geometry}$ are shown in Figure~\ref{fig:crab_obs}.
The selections of Energy, \texttt{OFFAXIS}, and $\theta_{geometry}$ are not applied in the histograms of
Energy, \texttt{OFFAXIS}, and $\theta_{geometry}$, respectively.
The red histograms are made from the events during the Crab GTI, and, the 
events collected during the period one day earlier than the Crab GTI are shown in black as a reference.

\begin{figure}[h]
 \begin{center}
  \includegraphics[width=0.4\textwidth]{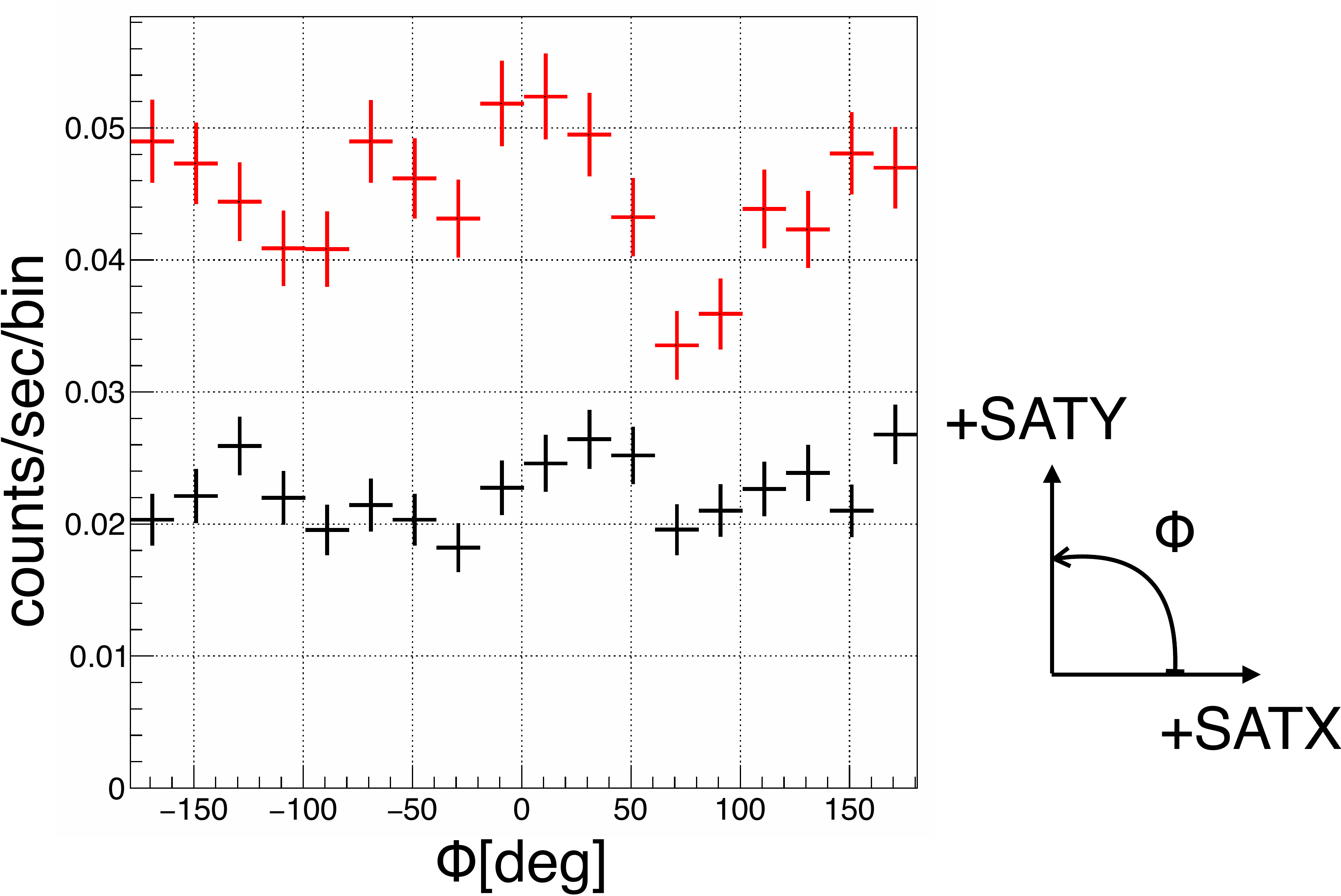} 
 \end{center}
\caption{The azimuth angle distributions obtained with the SGD Compton cameras.  
The red and the black show the distribution during the Crab GTI and that from an epoch 
one day earlier than the Crab GTI, respectively.
The definition of $\Phi$ is also shown. SATX and SATY mean the satellite $+$X-axis 
and the satellite $+$Y-axis, respectively.}\label{fig:crab_obs_phi}
\end{figure}

We measure the gamma-ray polarization by investigating the azimuth angle distribution in the Compton camera since gamma rays 
tend to be scattered perpendicular to the direction of the polarization vector of the incident gamma ray in Compton scatterings.
Figure~\ref{fig:crab_obs_phi} shows the azimuth angle 
distribution of Compton events obtained with the SGD Compton cameras.  The red and 
the black points show the distribution during the Crab GTI and that from 
one day earlier than the Crab GTI, respectively. The azimuthal angle $\Phi$ is defined as the 
angle from the satellite $+$X-axis to the satellite $+$Y-axis.
The average count rate during the Crab GTI is 0.808~count~s$^{-1}$.

\newpage

\subsection{Background Estimation for Polarization Analysis \label{sec:background}}

Before the Crab observations, Hitomi also observed RXJ~1856.5$-$3754 which is fairly faint in the energy band of the SGD (Hitomi Soft X-ray Imager results were reported in \cite{nakajima:2018}).
The GTIs of RXJ~1856.5$-$3754 and the exposure times are listed in 
Table~\ref{tab:rxjgti} and Table~\ref{tab:rxjexposure}, respectively.  The total exposure time 
of the all RXJ~1856.5$-$3754 observation is about 85.6~ksec and the number of the Compton-reconstructed events is about 24400. 
More than ten times larger number of events are available by using this 
observation than the observation of the Crab nebula.  
In order to obtain the azimuth angle distribution of the background events with 
better statics, the SGD data during the RXJ~1856.5$-$3754 GTI were investigated.

\begin{table*}
  \tbl{The good time intervals of the RXJ~1856.5$-$3754 observation. }{%
  \begin{tabular}{ccccc}
      \hline
      TSTART [s]\footnotemark[$\dag$] & TSTART [UTC] & TSTOP [s]\footnotemark[$\dag$] & TSTOP [UTC] & duration [s]\\ 
      \hline
	   70207640 & 2016/03/23 14:07:19 & 70212120 & 2016/03/23 15:21:59 & 4480 \\
	   70213720 & 2016/03/23 15:48:39 & 70218300 & 2016/03/23 17:04:59 & 4580 \\
	   70219740 & 2016/03/23 17:28:59 & 70221820 & 2016/03/23 18:03:39 & 2080 \\
	   70221860 & 2016/03/23 18:04:19 & 70224420 & 2016/03/23 18:46:59 & 2560 \\
	   70225700 & 2016/03/23 19:08:19 & 70230580 & 2016/03/23 20:29:39 & 4880 \\
	   70231600 & 2016/03/23 20:46:39 & 70236720 & 2016/03/23 22:11:59 & 5120 \\
	   70237100 & 2016/03/23 22:18:19 & 70274520 & 2016/03/24 08:41:59 & 37420 \\
	   70275720 & 2016/03/24 09:01:59 & 70280400 & 2016/03/24 10:19:59 & 4680 \\
	   70287960 & 2016/03/24 12:25:59 & 70292460 & 2016/03/24 13:40:59 & 4500 \\
	   70294120 & 2016/03/24 14:08:39 & 70298640 & 2016/03/24 15:23:59 & 4520 \\
	   70300140 & 2016/03/24 15:48:59 & 70304760 & 2016/03/24 17:05:59 & 4620 \\
	   70306140 & 2016/03/24 17:28:59 & 70310880 & 2016/03/24 18:47:59 & 4740 \\
	   70312120 & 2016/03/24 19:08:39 & 70317050 & 2016/03/24 20:30:49 & 4930 \\
	   70317950 & 2016/03/24 20:45:49 & 70355100 & 2016/03/25 07:04:59 & 37150 \\
      \hline
    \end{tabular}}\label{tab:rxjgti}
\begin{tabnote}
\footnotemark[$\dag$] : The unit for TSTART and TSTOP is AHTIME.
\end{tabnote}
\end{table*}

\begin{table}
  \tbl{Exposures of the RXJ~1856.5$-$3754 observation.}{%
  \begin{tabular}{cc}
      \hline
       & Live Time         \\
      \hline
      SGD1 CC1 & 84358.5  \\
      SGD1 CC2 & 84432.5  \\
      SGD1 CC3 & 84559.5  \\
      SGD2 CC1 & 89159.2  \\
      \hline
    \end{tabular}}\label{tab:rxjexposure}
\begin{tabnote}
\end{tabnote}
\end{table}

\begin{figure*}[h]
 \begin{center}
  \includegraphics[width=6cm]{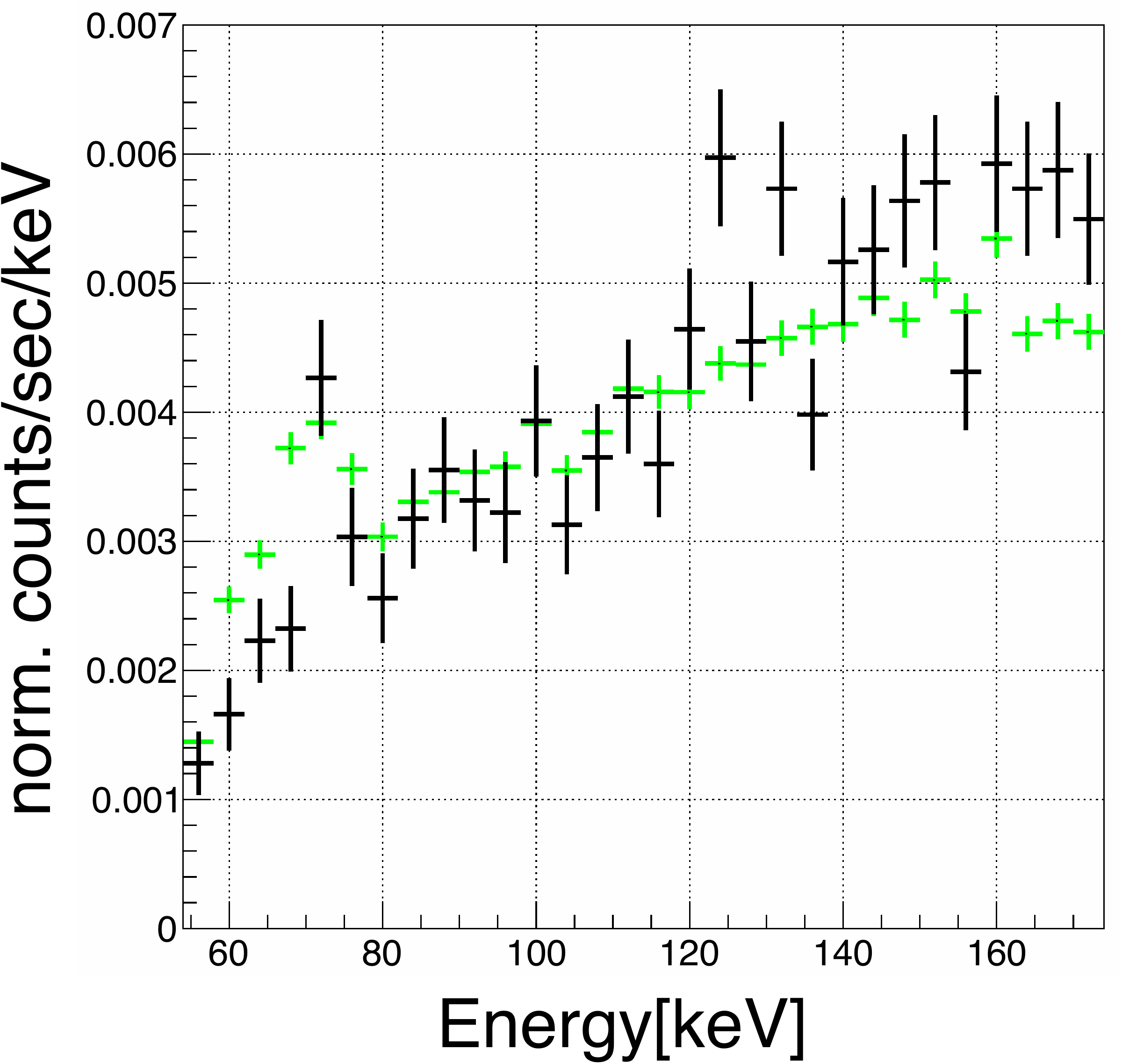} 
  \includegraphics[width=6cm]{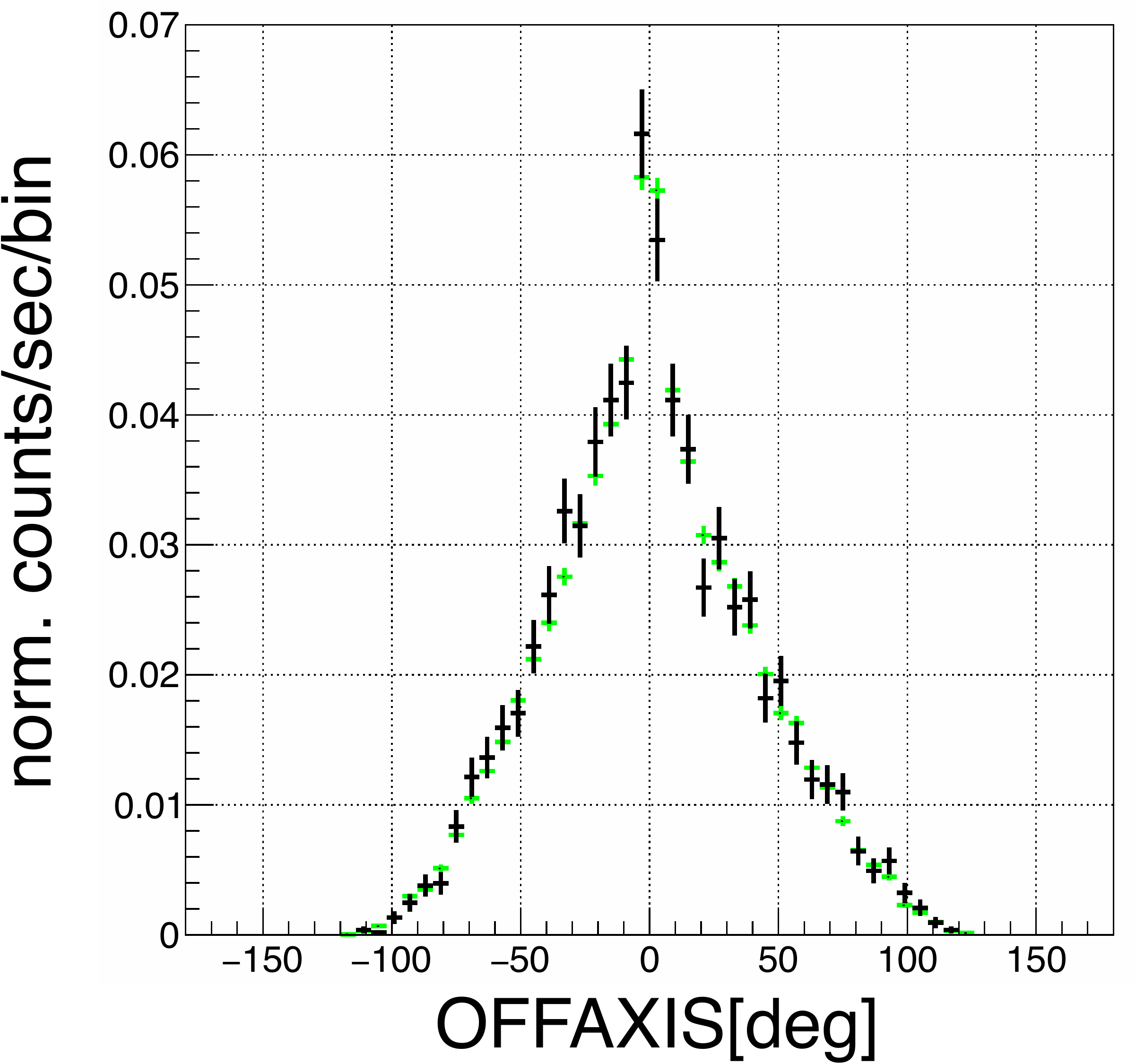} 
  \\
  \bigskip
  \includegraphics[width=6cm]{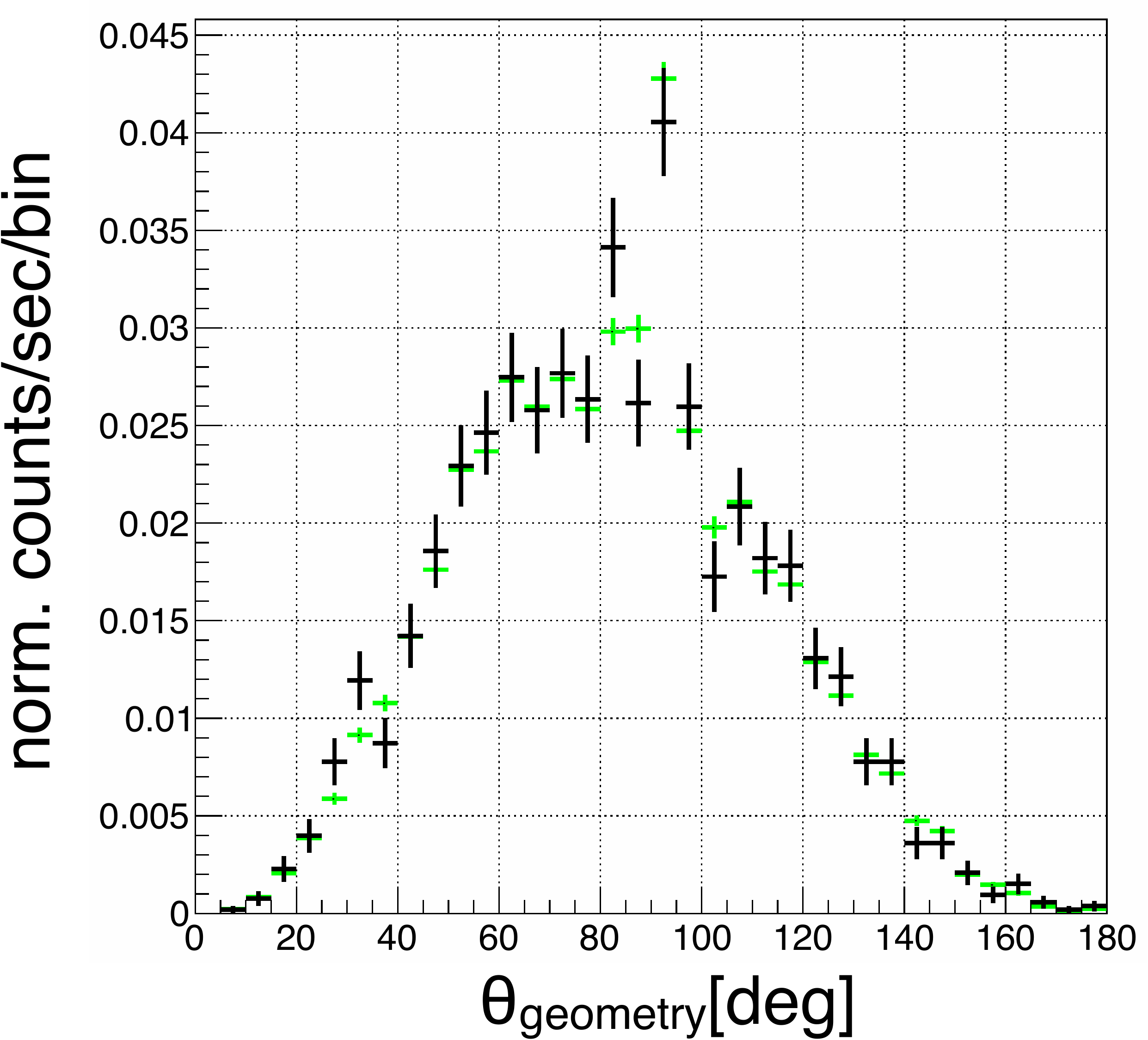} 
  \includegraphics[width=6cm]{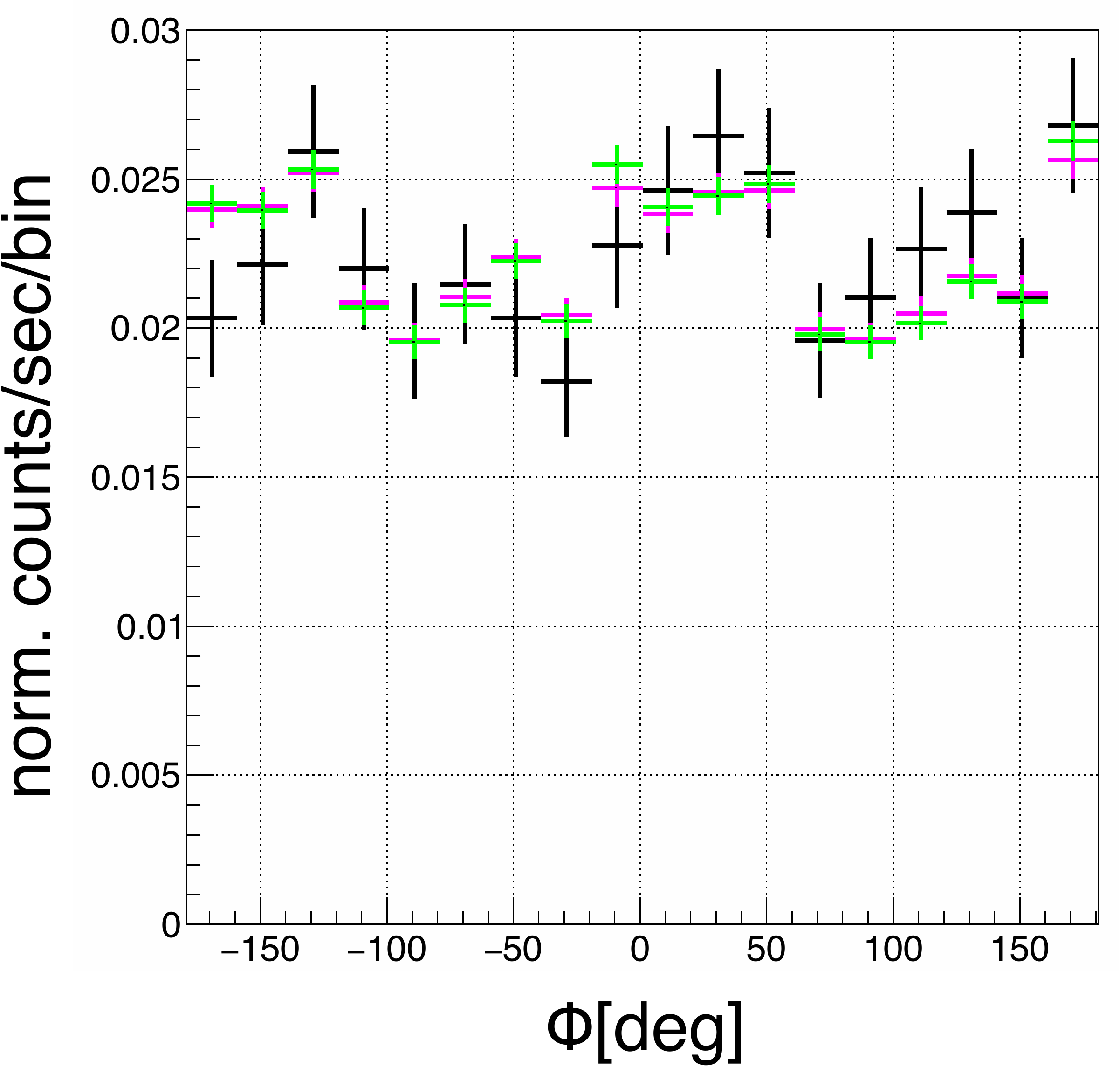} 
  \\
  \bigskip
 \end{center}
\caption{The comparisons between the all RXJ~1856.5$-$3754 observation and those obtained 
one day earlier than the Crab GTI. The green and the black show the all RXJ~1856.5$-$3754 observation data and 
one day earlier than the Crab GTI data, respectively. The black data points are identical to the black points 
in Figure~\ref{fig:crab_obs} and Figure~\ref{fig:crab_obs_phi}.
The normalizations of the histograms for the all RXJ~1856.5$-$3754 observation are scaled to match with 
the count rate of the one day earlier Crab GTI.}\label{fig:compare_bgds}
\end{figure*}

Comparisons of the incident energy, \texttt{OFFAXIS}, $\theta_{geometry}$ and the azimuth angle $\Phi$ between the RXJ~1856.5$-$3754 GTI 
and one day earlier than the Crab GTI are shown in Figure~\ref{fig:compare_bgds}.
Since orbits with no SAA passage are included in the all RXJ~1856.5$-$3754 
observation, the flux level was lower than that obtained one day earlier than 
the Crab GTI.  The count rate of the events during 
the RXJ~1856.5$-$3754 GTI is 0.285~count~s$^{-1}$, and that during the one day 
earlier than Crab GTI is 0.404~count~s$^{-1}$.  Therefore, the scale of the histograms 
for the RXJ~1856.5$-$3754 GTI are normalized to match those for one day earlier than the Crab GTI. 
The distributions of \texttt{OFFAXIS}, $\theta_{geometry}$ and the azimuth angle $\Phi$ are similar.
Since the incident energy spectrum of the RXJ~1856.5$-$3754 GTI looks slightly different from that observed one day earlier than the Crab GTI, we further investigated the effect on the $\Phi$ distrbution.
We divided the data in five energy bands, 60--80~keV, 80--100~keV, 100--120~keV, 120--140~keV and 140--160~keV, and the number of events in each energy band is normalized to match those for one day earlier than the Crab GTI.
The resulting $\Phi$ distribution for the RXJ~1856.5$-$3754 GTI is shown as the magenta points in the lower right panel of Figure~\ref{fig:compare_bgds}.
We do not observe any significant trend from the original distribution for the RXJ~1856.5$-$3754 GTI, 
which implies that the difference in the energy spectrum does not have significant effect on the $\Phi$ distribution.
From above investigations, we conclude that the Compton reconstructed events during the RXJ~1856.5$-$3754 GTI can be utilized for the background estimation of the polarization analysis.

\newpage

\subsection{Monte Carlo simulation}

Monte Carlo simulations of SGD are essential to derive physical parameters including gamma-ray polarization 
from the observation data.
For the Monte Carlo simulations, we used \textit{ComptonSoft}\footnote{https://github.com/odakahirokazu/ComptonSoft} 
in combination with a mass model of the SGD and databases describing detector parameters that affect 
the detector response to polarized gamma rays.  \textit{ComptonSoft} is a general-purpose 
simulation and analysis software suite for semiconductor radiation detectors including 
Compton cameras \citep{Odaka:2010}, and depends on the \textsc{Geant4} toolkit library 
\citep{Agostinelli:2003, Allison:2006, Allison:2016} for the Monte Carlo simulation of 
gamma rays and their associated particles.
We chose \textsc{Geant4} version 10.03.p03 and \texttt{G4\-Em\-Livermore\-Polarized\-Physics} 
as the physics model of electromagnetic processes.  
The mass model describes the entire structure of one SGD unit including the surrounding BGO shields.
The databases of the detector parameters contain configuration of readout electrodes, 
charge collection efficiencies, energy resolutions, trigger properties, and data 
readout thresholds in order to obtain accurate detector responses of the semiconductor 
detectors and scintillators composing the SGD unit.

The format of the simulation output file is same as the SGD observation data. 
The simulation data can be processed with \texttt{sgdevtid}, and as a result, it is guaranteed 
that the same event reconstructions are performed for both observation data and simulation data.
 
For the Compton camera part, the accuracy of the simulation response to the gamma-ray 
photons and the gamma-ray polarization was confirmed through polarized gamma-ray beam 
experiments performed at SPring-8 \citep{Katsuta:2016}.  
The better than 3\% systematic uncertainty was validated in the polarized gamma-ray beam 
experiments.
On the other hand, the effective 
area losses due to the distortions and misalignments of the fine collimators (FCs) 
are not implemented in the SGD simulator \citep{Tajima:2018}.  We have not obtained measurements 
of the FC distortions and misalignments with the calibration observations:  this 
is because the satellite operation was terminated before we had opportunities to 
make such measurements.  In the simulator, the ideal shape FCs with no distortion 
and no misalignment are implemented. 
Since the losses due to the distortions and misalignments of the fine collimators does not affect 
the azimuthal angle distribution of the Compton scattering, 
the effects on the polarization measurements are negligible.

\begin{figure}[h]
 \begin{center}
  \includegraphics[width=0.4\textwidth]{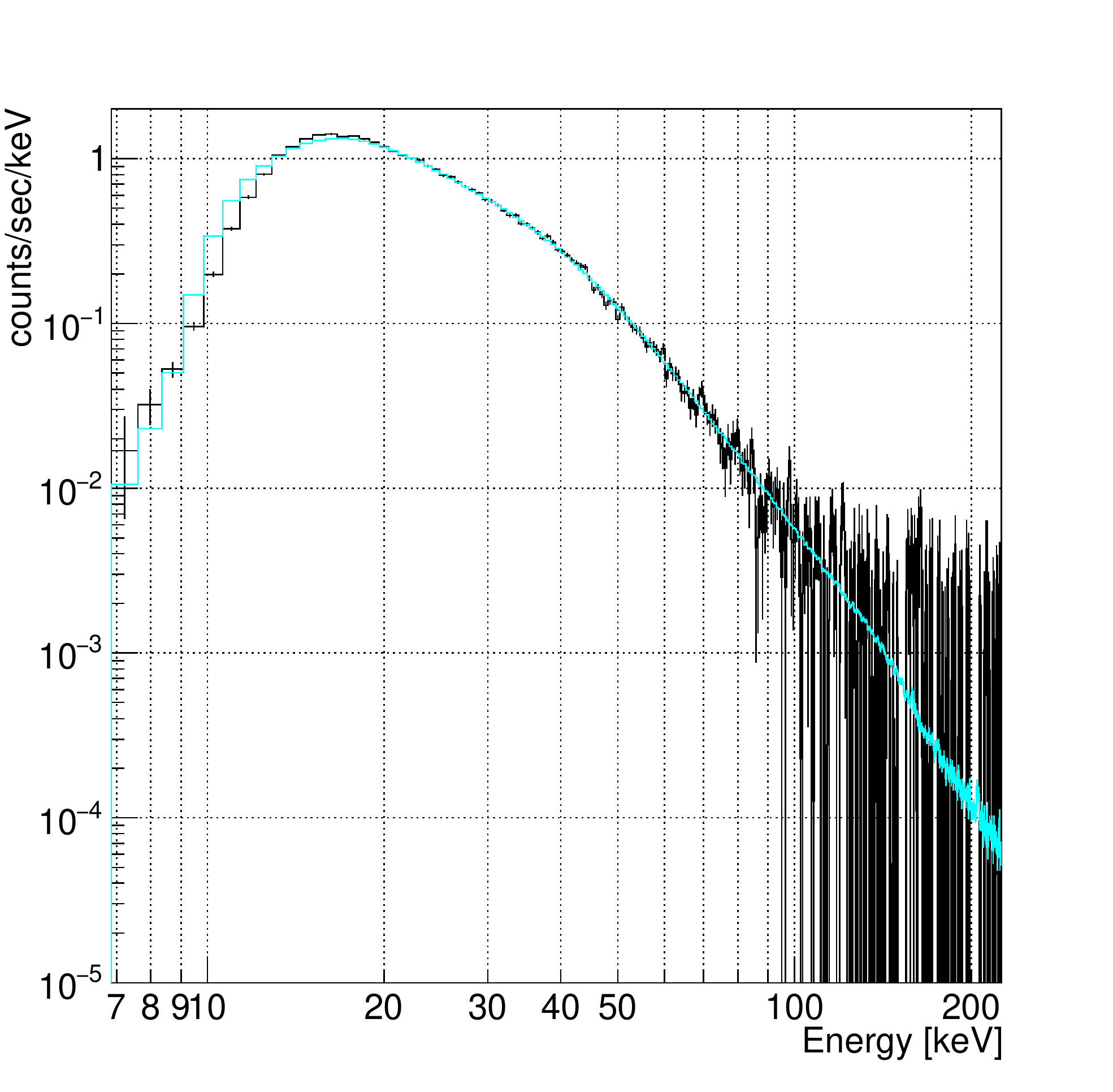} 
 \end{center}
\caption{The single hit spectra in the Si detectors of the Crab observation obtained 
with the four Compton cameras. The background subtracted observation spectrum
is shown in black, and the simulation spectrum is shown in cyan. In the simulation, 
a power-law spectrum, $N \cdot (E/1~\mathrm{keV})^{-\Gamma}$ is assumed, 
with a photon index ($\Gamma$) of 2.1 and $N$~=~8.23.}\label{fig:si_spec}
\end{figure}

In the simulation of the Crab nebula emission, we assumed a power-law spectrum, 
$N \cdot (E/1~\mathrm{keV})^{-\Gamma}$, with a photon index ($\Gamma$) of 2.1.
In the first step, unpolarized gamma-ray photons are assumed.
And, the normalization of the simulation model ($N$) is derived from the Si single-hit events.
Figure~\ref{fig:si_spec} shows the Si single hit spectrum obtained with the 4 Compton cameras. 
The background spectrum is estimated from the observations taken one day earlier than those for the 
Crab GTI, and the background subtracted spectrum is shown in Figure~\ref{fig:si_spec}, together with   
the simulated spectrum.  By scaling the 
integrated rate of the simulation spectrum in the 20--70~keV range to match the observed 
rate, we obtained $N$~=~8.23 which corresponds to a flux of 1.89~$\times$~10$^{-8}$~erg/s/cm$^{2}$ 
in the 2--10~keV energy range.

\begin{figure}[h]
 \begin{center}
  \includegraphics[width=0.4\textwidth]{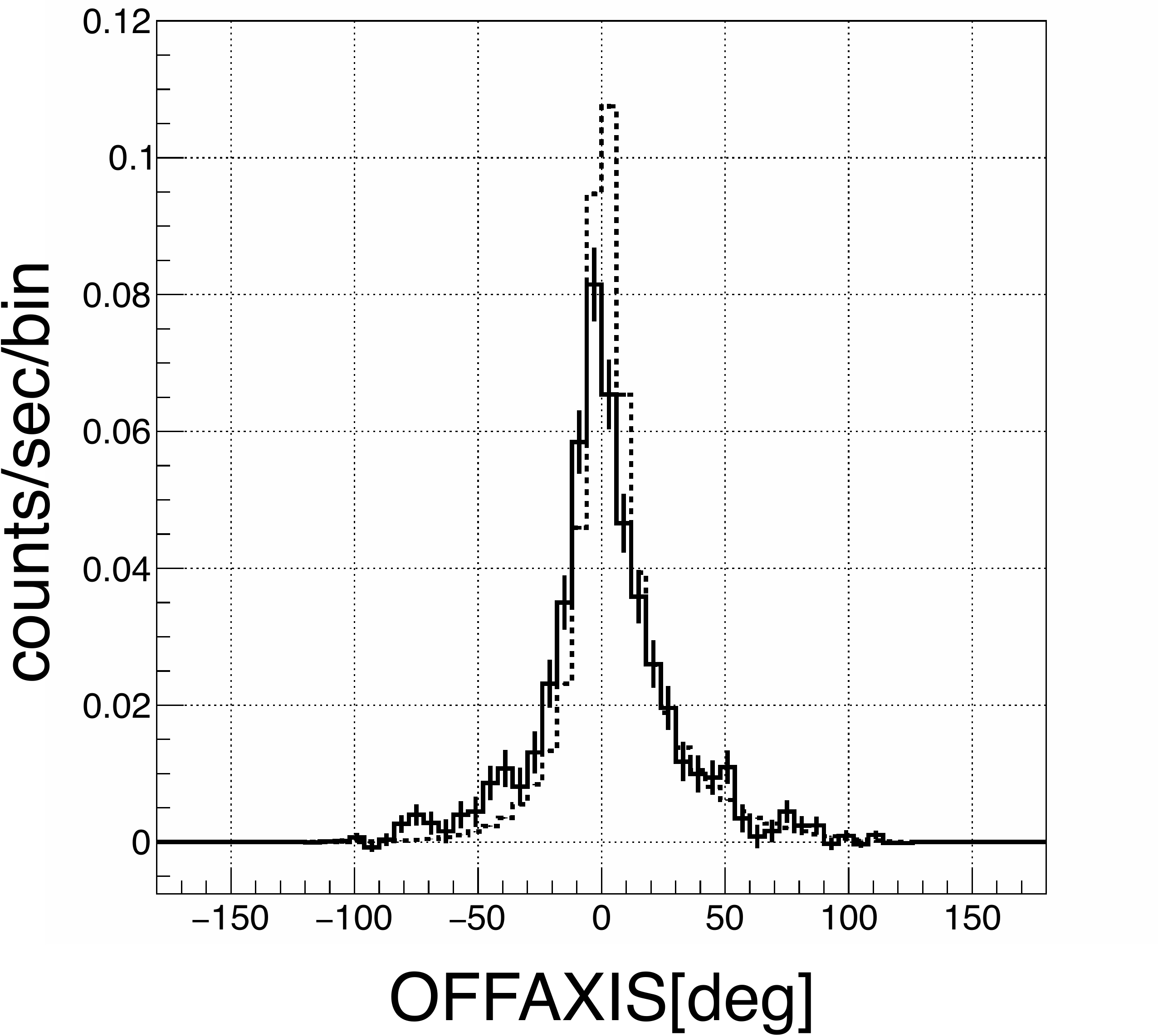} 
 \end{center}
\caption{The comparisons of the distributions of \texttt{OFFAXIS} events 
between the observation and the simulation.  The solid line and the dotted line 
show the observation data and the simulation data, respectively.}\label{fig:obssim_arm}
\end{figure}

\begin{figure}[h]
 \begin{center}
  \includegraphics[width=0.4\textwidth]{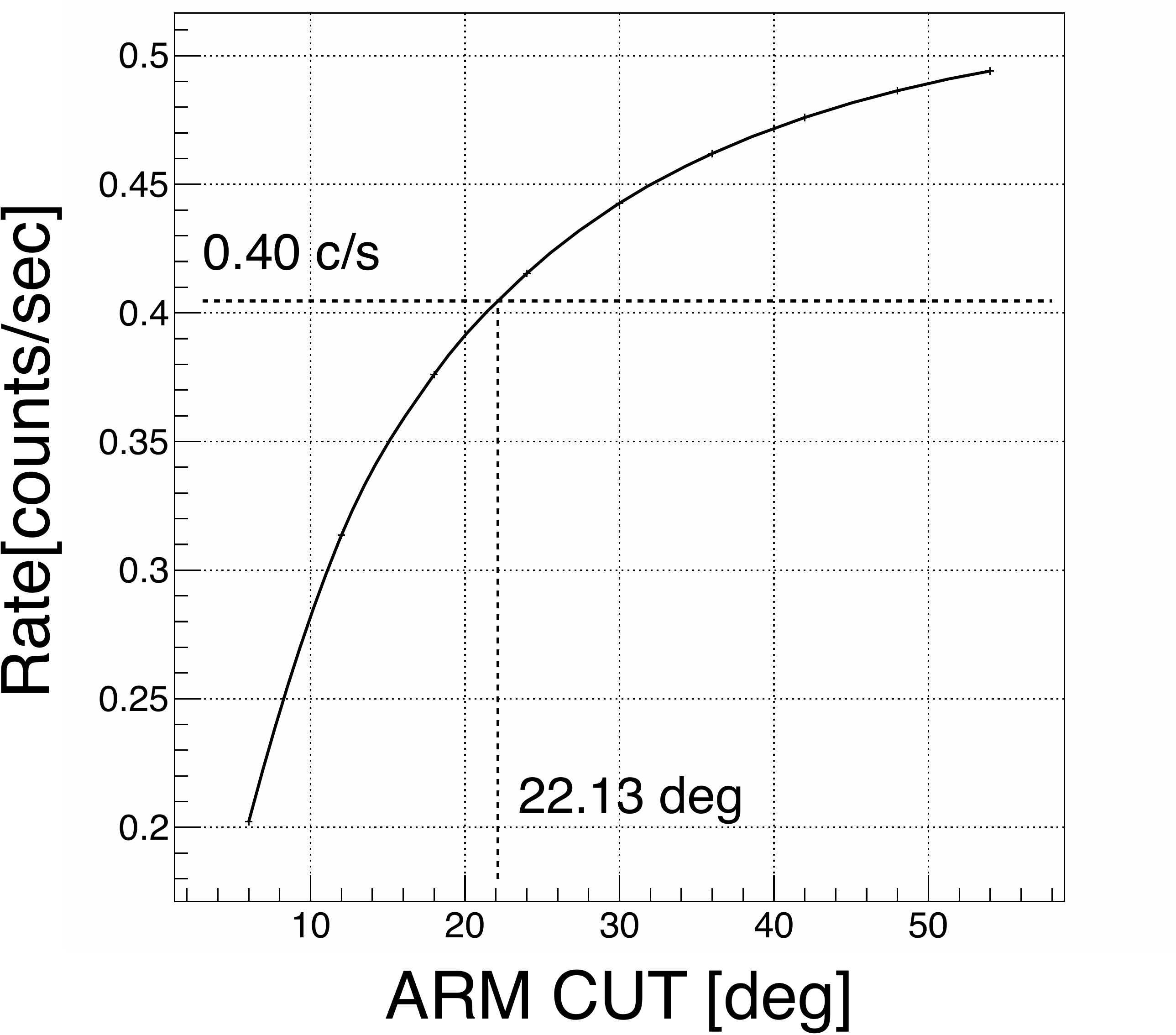} 
 \end{center}
\caption{The relation between the count rate and the \texttt{OFFAXIS} selection for 
the simulation events. The count rate of 0.40~count~s$^{-1}$ derived from the observation data corresponds 
to the \texttt{OFFAXIS} selection of 22.13\degree~.}\label{fig:sim_armcut}
\end{figure}

We compare the Compton reconstructed events between 
the observation and the simulation.  In Figure~\ref{fig:obssim_arm}, the distributions 
of \texttt{OFFAXIS} for the observation and the simulation are shown.  The distribution of \texttt{OFFAXIS} 
for the simulation is slightly narrower than that for the observation. If the same selection 
of $-$30\degree~$<$~\texttt{OFFAXIS}~$<$~$+$30\degree~ is applied for the both events, the observation 
count rate becomes 8.6\% smaller than the simulation count rate. 
We think that one of causes of this discrepancy is in modeling the Doppler broadening profile of Compton scattering for 
electrons in silicon crystals. However, at this time, we have not found a solution to eliminate the discrepancy from first principles.
Therefore, by adjusting the \texttt{OFFAXIS} selection value of the simulation, we decided to match the count rate of the simulation 
to the observed count rate of 0.40 count sec$^{-1}$.
The relation between the count rate and the \texttt{OFFAXIS} selection for the simulation events is shown in Figure~\ref{fig:sim_armcut}.
From the relation, we obtained 22.13~degree as the \texttt{OFFAXIS} selection value of the simulation.
The effect of adjusting the \texttt{OFFAXIS} selection for the simulation is discussed later in this section.

\begin{figure*}[h]
 \begin{center}
  \includegraphics[width=6cm]{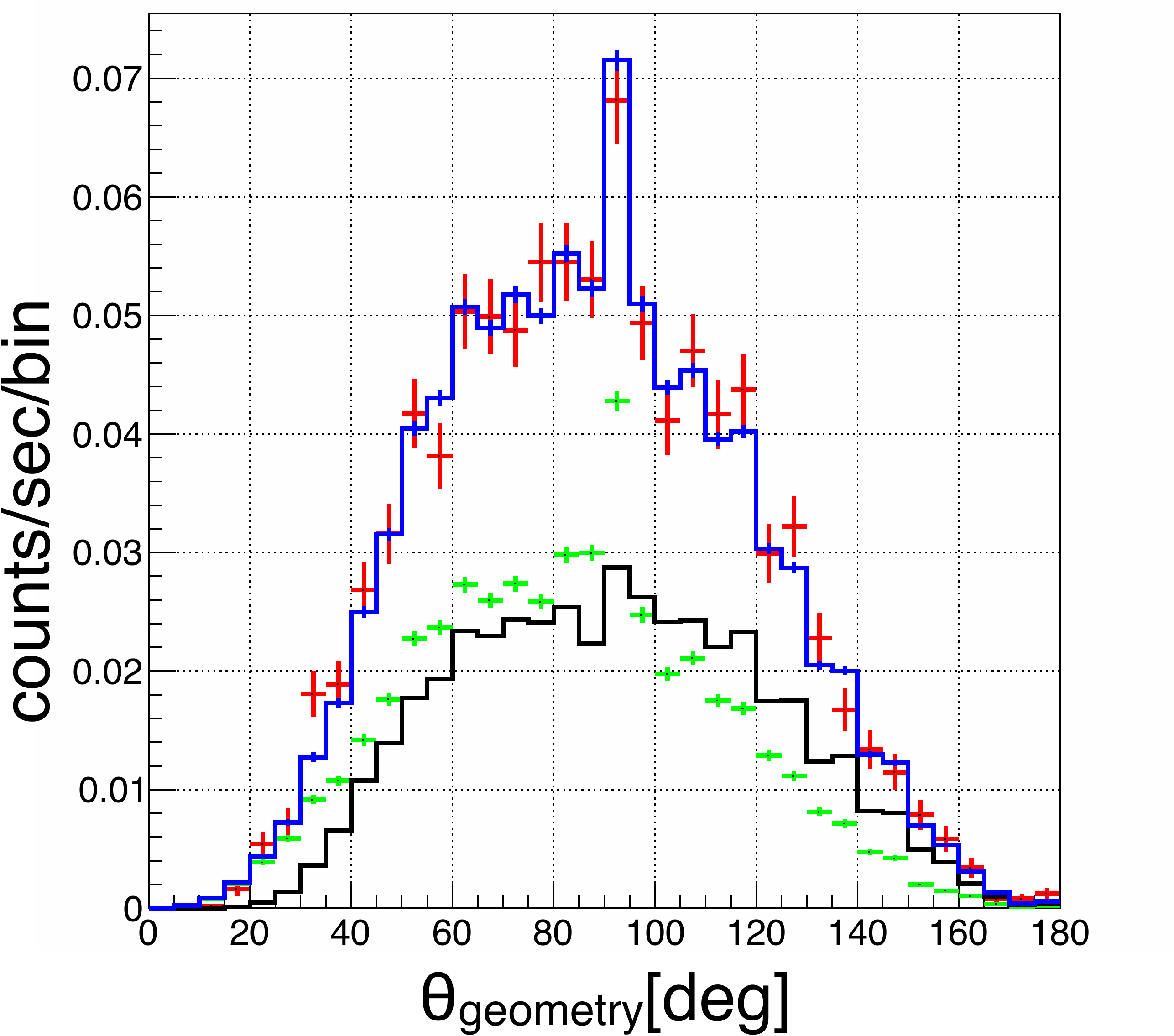} 
  \includegraphics[width=6cm]{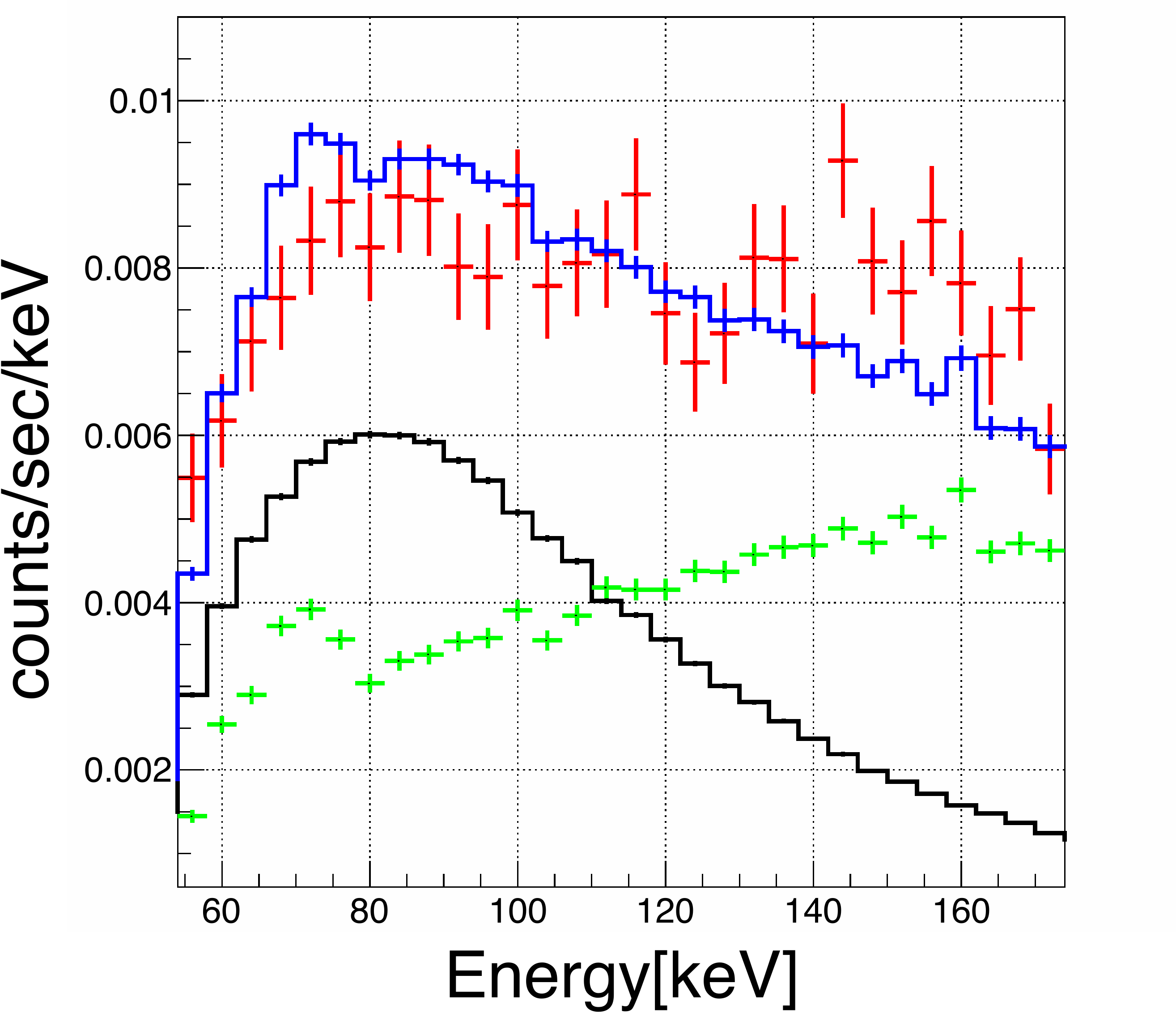} 
 \end{center}
\caption{The distribution of the $\theta_{\rm geometry}$ (left) and the energy spectrum (right). 
The observational data are plotted in red.
The simulation data with the selection of $-$22.13\degree~$<$~\texttt{OFFAXIS}~$<$~$+$22.13\degree~ 
are shown in black, and, 
the background data derived from the all RXJ~1856.5$-$3754 observation are shown in green.
Sum of the simulation data and the background data are plotted in blue. The red data points are identical to the red ones in Figure~\ref{fig:crab_obs}, 
and, the green data points are identical to the green ones in Figure~\ref{fig:compare_bgds}.
}\label{fig:compare_obs_sim}
\end{figure*}

The observational data, the background data, and the simulation data are plotted in Figure~\ref{fig:compare_obs_sim}.
The simulation data with the selection of 
$-$22.13\degree~$<$~\texttt{OFFAXIS}~$<$~$+$22.13\degree~ is shown in black, the background data 
derived from the entire RXJ~1856.5$-$3754 observation is shown in green.  
Sum of the simulation data and the background data is plotted in blue, and, 
is comparable with the observation data shown in red.  
The $\theta_{\rm geometry}$ distribution is well reproduced by the Monte Carlo simulation 
while the energy spectrum shows small discrepancy due to the background data as shown in section~\ref{sec:background}.

\begin{figure*}[h]
 \begin{center}
  \includegraphics[width=6cm]{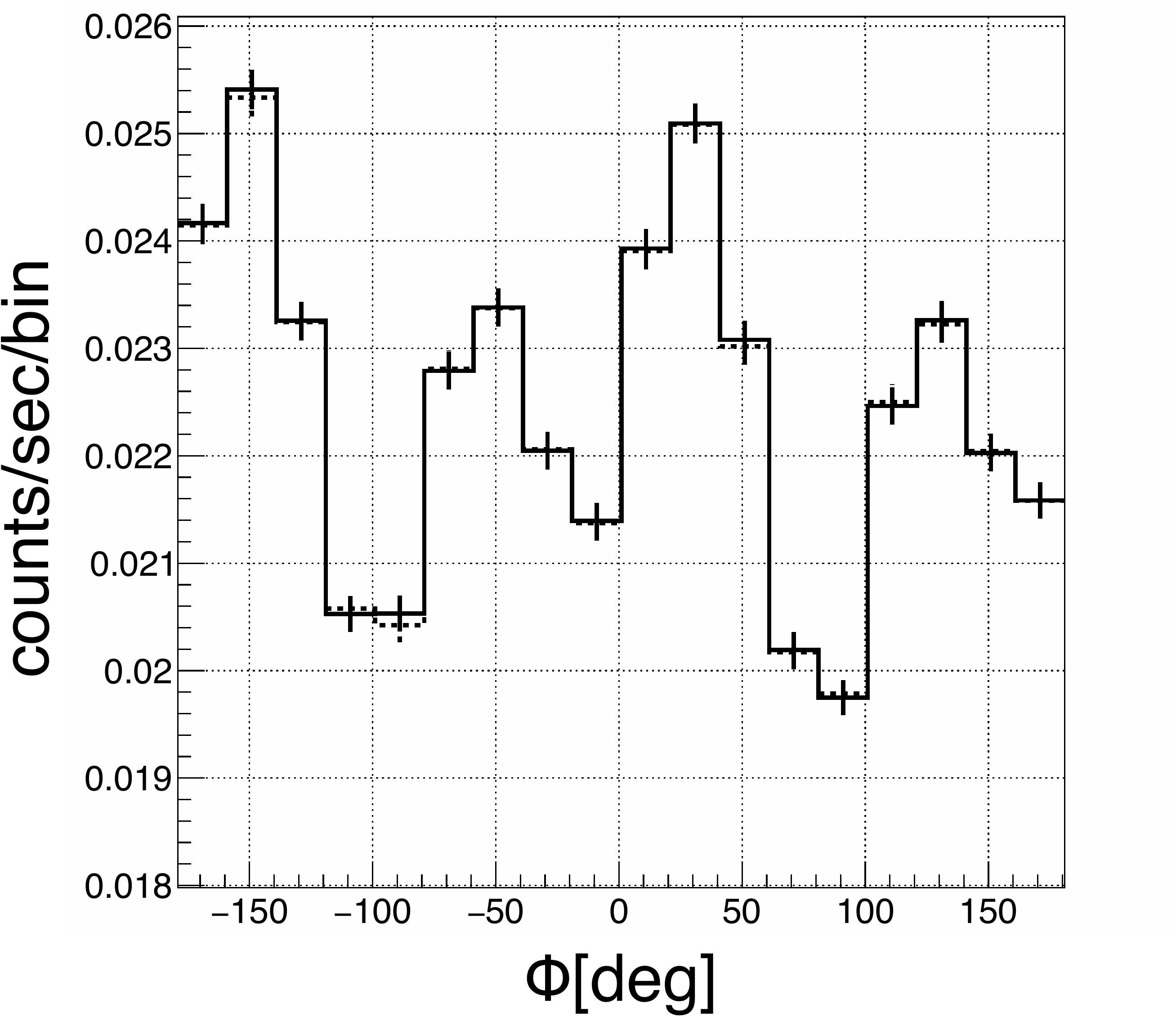} 
  \includegraphics[width=6cm]{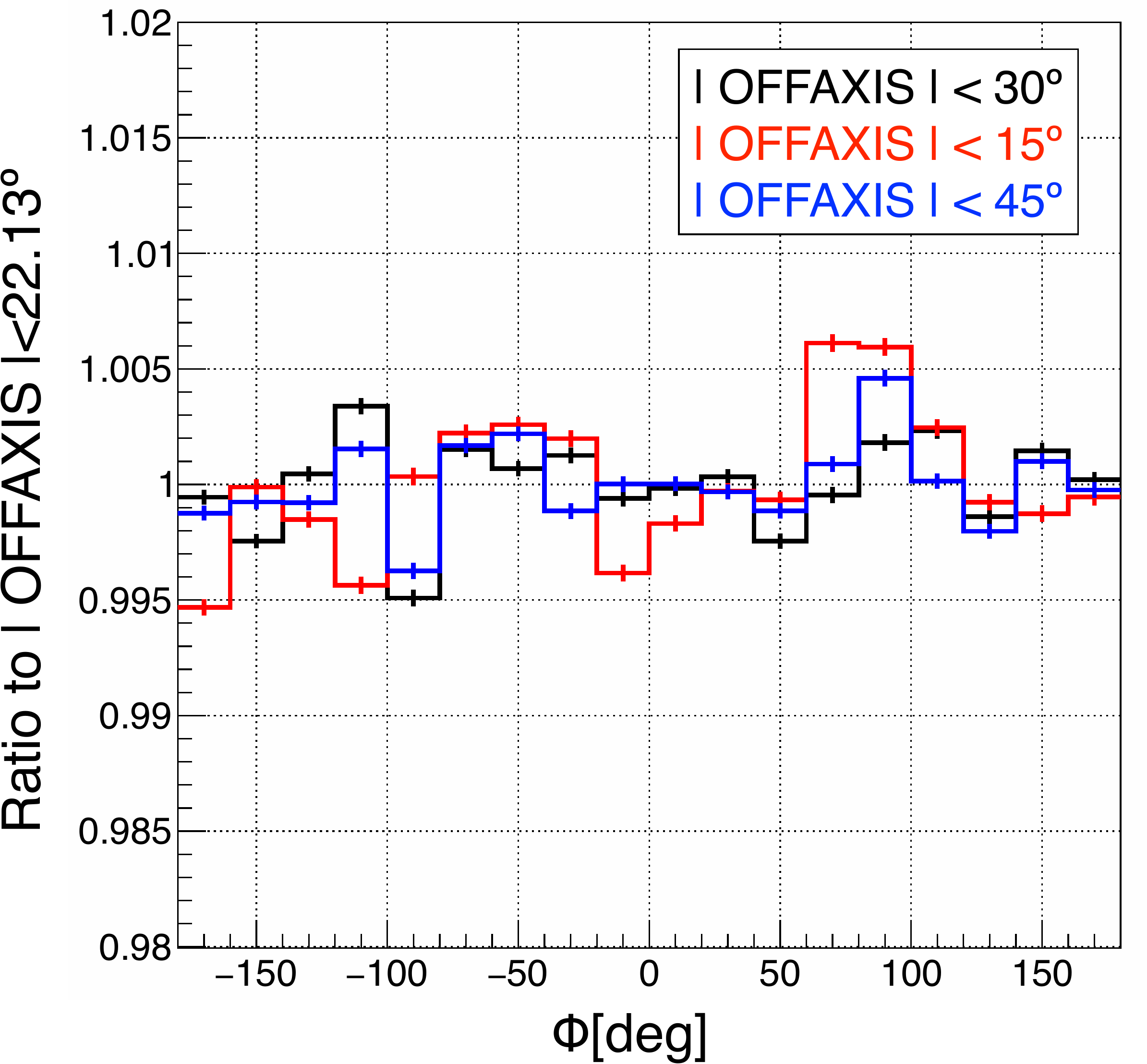} 
 \end{center}
\caption{The azimuth angle distributions of simulation data. (left): 
The azimuth angle distributions of the simulation data with the \texttt{OFFAXIS} selection of 
$-$22.13\degree~$<$~\texttt{OFFAXIS}~$<$~$+$22.13\degree~ and $-$30\degree~$<$~\texttt{OFFAXIS}~$<$~$+$30\degree~ 
are shown in the solid line and the dotted line, respectively.
The normalization for the $-$30\degree~$<$~\texttt{OFFAXIS}~$<$~$+$30\degree~ selection is scaled. 
(right): The dependence of the azimuth angle distribution on the selected \texttt{OFFAXIS} value.  
This is shown as the ratio to the \texttt{OFFAXIS} selection of three values to that limited to 
$-$22.13\degree~$<$~\texttt{OFFAXIS}~$<$~$+$22.13\degree~.  The black, red and blue show the results 
for \texttt{OFFAXIS} selections of $-$30\degree~$<$~\texttt{OFFAXIS}~$<$~$+$30\degree~, 
$-$15\degree~$<$~\texttt{OFFAXIS}~$<$~$+$15\degree~, and $-$45\degree~$<$~\texttt{OFFAXIS}~$<$~$+$45\degree~, respectively.
}\label{fig:sim_hist_phi}
\end{figure*}

The azimuth angle distributions of the simulated data are shown in Figure~\ref{fig:sim_hist_phi}.
The left-hand side panel shows the azimuth angle distributions of the simulated data with the 
\texttt{OFFAXIS} selection of $-$22.13\degree~$<$~\texttt{OFFAXIS}~$<$~$+$22.13\degree~ and 
$-$30\degree~$<$~\texttt{OFFAXIS}~$<$~$+$30\degree~.  The normalization for the 
$-$30\degree~$<$~\texttt{OFFAXIS}~$<$~$+$30\degree~ selection is scaled.  
There is little difference in the azimuth angle distribution between these two selections.
The right-hand side panel of Figure~\ref{fig:sim_hist_phi} shows how the 
azimuth angle distribution depends on the \texttt{OFFAXIS} selection. 
It is found that 
the azimuth angle distribution changes by less than 1\% when the angle selection range is changed from 15\degree\ to 45\degree.

\newpage

\section{Polarization Analysis}\label{sec:polarization_analysis}
\subsection{Parameter search for the polarization measurement}

\begin{figure}
\begin{center}
\includegraphics[width=0.45\textwidth]{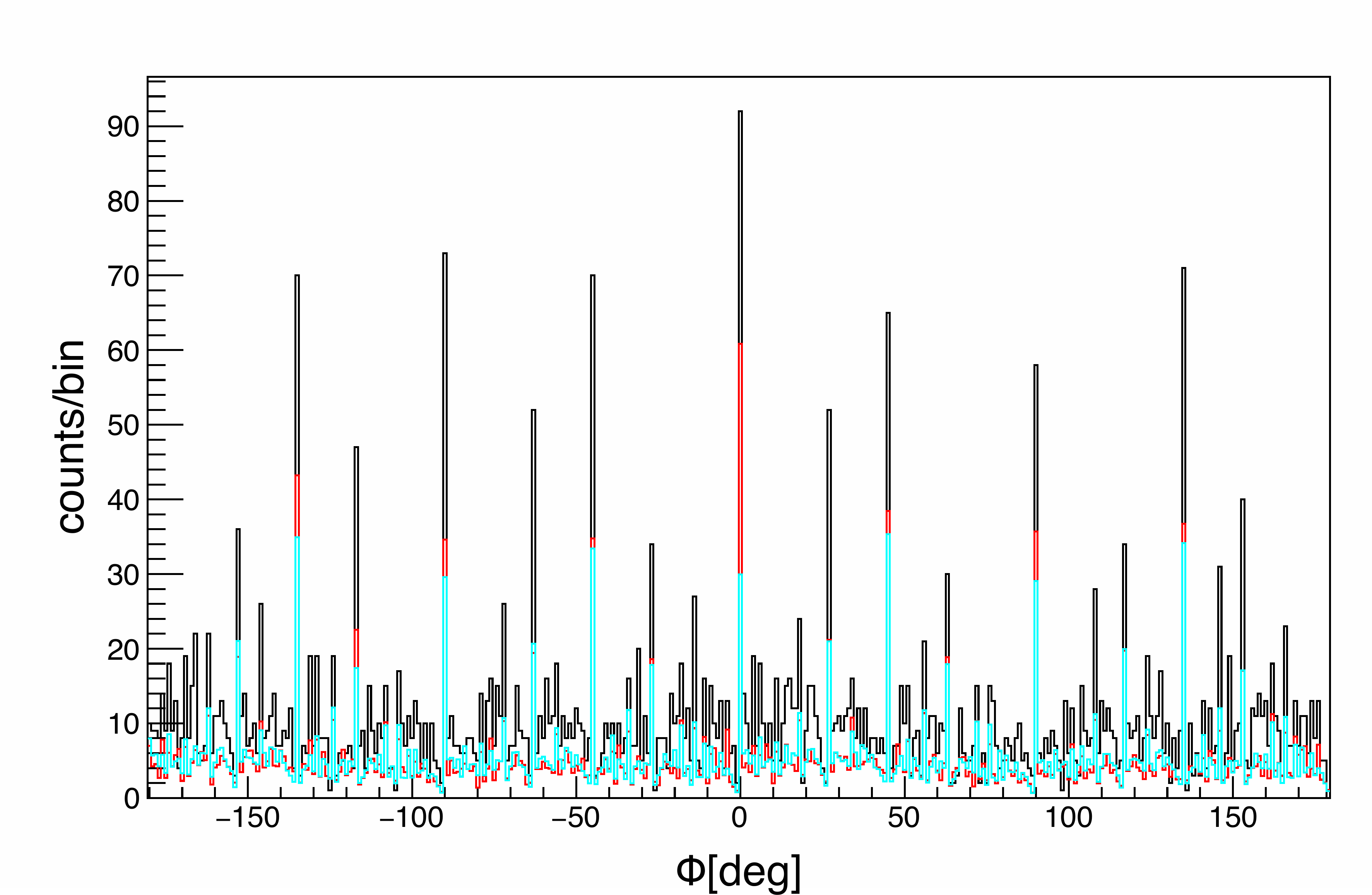}
\end{center}
\caption{The distributions of the azimuthal angle. The black, red and cyan lines show the 
Crab GTI data, the background data derived from all the RXJ~1856.5$-$3754 observation, 
and simulation data for unpolarized gamma-rays, respectively. Each exposure time is matched with 
the exposure time of the observation during the Crab GTI. The bin width is 1 degree.
}
\label{fig:obs_vs_bkgrxj_azimuth}
\end{figure}

We obtained the azimuth angle distributions for the Crab observation, the background, 
and the unpolarized gamma-ray simulation, respectively.  
In order to derive the polarization 
parameters of the Crab nebula from these data, we adopt a binned likelihood fit.
Although the bin width of the histograms for the azimuth angle distributions 
was 20~\degree for the figures in the previous subsections, 
1~\degree per bin histograms are prepared for the binned likelihood fit. 
The histograms are shown in Figure~\ref{fig:obs_vs_bkgrxj_azimuth}. For the simulation data, 
the \texttt{OFFAXIS} selection of $-$22.13\degree~$<$~\texttt{OFFAXIS}~$<$~$+$22.13\degree~ is adopted.

In the binned likelihood fit, we scaled the background data 
and unpolarized simulation with the exposure time of the Crab GTI.
Expected counts $n_\mathrm{exp} (\phi_i)$ in each bin are expressed by the following 
equation using the background ${n_{\mathrm{bkg}}}(\phi_i)$ and unpolarized simulation 
data $n_\mathrm{sim}(\phi_i)$ in count space:
\begin{equation}
\label{eq:n_exp}
n_{\mathrm{exp}}(\phi_i) = n_\mathrm{sim}\left(\phi_i \right) \left(1 - Q \cos\left(2\left(\phi_i - \phi_0\right)\right)\right) + n_\mathrm{bkg}(\phi_i), 
\end{equation}
where $Q$ is a modulation amplitude due to a polarization, $\phi_0$ is a polarization angle in the coordinate of the Compton camera, $i$ is a bin number ($i>=1$), and $\phi_i$ is the azimuthal angle at $i$-th bin center.
We assume that the Crab observation counts $n_\mathrm{obs}$ is given by 
Poisson distributions which can be expressed as
\begin{equation}
\label{eq:poisson}
\mathrm{Poisson}\left(n_\mathrm{obs}(\phi_i) | n_\mathrm{exp}(\phi_i)\right) = \frac{n_\mathrm{exp}^{n_\mathrm{obs}} e^{-n_\mathrm{exp}}}{n_\mathrm{obs} !}.
\end{equation}
Likelihood function is a product of the Poisson distributions:
\begin{equation}
\label{eq:likelihood}
L(\phi_0, Q) = \prod_i \mathrm{Poisson}(n_\mathrm{obs}(\phi_i) | n_\mathrm{exp} (\phi_i)).
\end{equation}
Best fit parameters of $Q$ and $\phi_0$, can be obtained by searching a combination of the parameters that yields the minimum of  
\begin{equation}
\mathcal{L} = -2\log L.
\end{equation}

The errors of estimated value are evaluated from the confidence level.
In the large data sample limit, the difference of the log likelihood $\mathcal{L}$ from the minimum $\mathcal{L}_0$, $\Delta \mathcal{L}=\mathcal{L}-\mathcal{L}_0$, follows $\chi^2$.
Since we have two free parameters, $\Delta \mathcal{L}$s of 2.30, 5.99, 9.21 correspond 
to the coverage probabilities of 68.3\%, 95.0\% and 99.0\%, respectively.

\subsection{Polarization results and validation}

\begin{figure}[h]
 \begin{center}
  \includegraphics[width=0.45\textwidth]{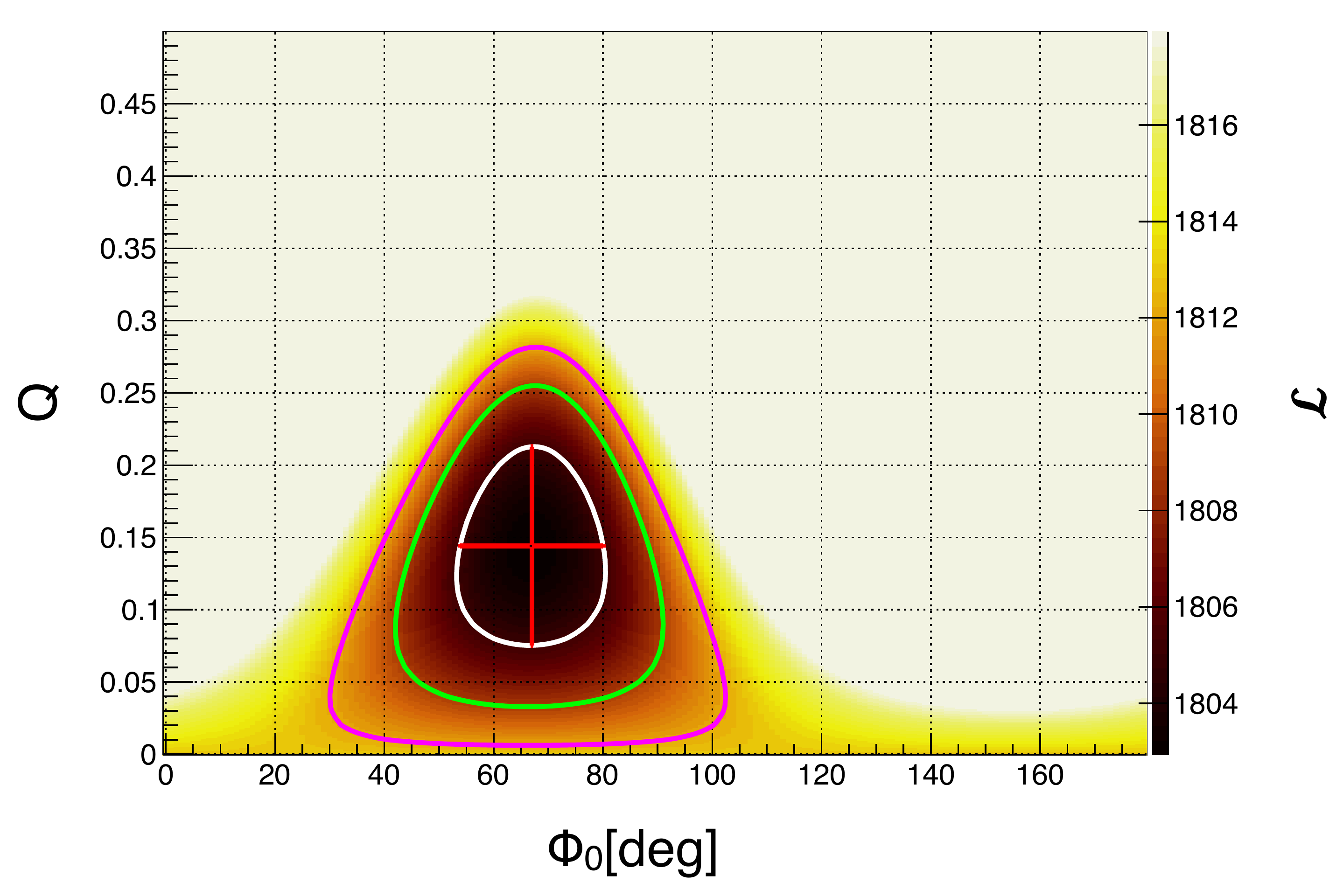} 
 \end{center}
\caption{The results of the maximum log likelihood estimation for the 
Crab observation data. The best fit parameters are shown in a red cross. 
The contours of $\Delta \mathcal{L}$s of 2.30, 5.99, and 9.21 are shown as 
the white line, the green line, and the magenta line, respectively.  
In the large sample limit, they correspond to the 
coverage probabilities of 68.3\%, 95.0\% and 99.0\%, respectively.
The best fit parameters are $Q = 0.1441$ and $\phi_0 =  67.02\degree$. 
The errors corresponding to $\Delta \mathcal{L} = 2.30$ level ($1\sigma$) are $-0.069, +0.069$ 
and $-13.2\degree\ , +13.0\degree$\ for $Q$ and $\phi_0$, respectively.}
\label{fig:crab_obs_result}
\end{figure}

The dependence of $\mathcal{L}$ on $Q$ and $\phi_0$ is shown in Figure~\ref{fig:crab_obs_result}. 
The best fit parameters of $Q$ and $\phi_0$, $Q = 0.1441$ and $\phi_0 =  67.02\degree$.
The contours in Figure~\ref{fig:crab_obs_result} show 
$\Delta \mathcal{L} = 2.30, 5.99, 9.21$. 
The errors corresponding to the $\Delta \mathcal{L} = 2.30$ level ($1\sigma$)
are $-0.0688, +0.0688$ and $-13.15\degree, +13.02\degree$ for $Q$ and $\phi_0$, respectively.
We also derived the log likelihood at $Q = 0$ ($\mathcal{L}_{Q=0}$), and then, 
the difference between $\mathcal{L}_{Q=0}$ and $\mathcal{L}_0$ is found to be 10.03.

The modulation amplitude for the 100\% polarized gamma-ray photons ($Q_{100}$) is slightly dependent on $\phi_0$ 
and is estimated to be $Q_{100} = 0.6534$ with the Monte Carlo simulation for $\phi_0 = 67\degree$ and a power-law spectrum 
with a photon index ($\Gamma$) of 2.1
As the result, the polarization fraction ($\Pi$) of the Crab nebula is calculated 
as $\Pi = 0.1441/0.6534 = 22.1\%$, and, the error is also calculated as $0.0688/0.6534 = 10.5\%$.

In order to validate the statistical confidence, we made 1000 simulated Crab 
observation data sets and derived the parameters with the binned likelihood fits for each data set.
Because the exposure time of the Crab observation was about 5~ksec, the exposure time 
of the simulated Crab observation data is also set to be 5~ksec.  
In the Monte Carlo simulations, 
the polarization fraction $\Pi = 0.22$ and the polarization angle $\phi_0 = 67\degree$\  are assumed. 
The background data is prepared using the azimuth angle distribution of the background 
data shown in Figure~\ref{fig:obs_vs_bkgrxj_azimuth}.  By using the random number according 
to the azimuth angle distribution of the background data, 1000 sets of 5~ksec background 
data are obtained.  The 1000 sets of the simulated Crab observation data are prepared by summing each Monte 
Carlo data and background data. 

\begin{figure}[ht]
 \begin{center}
  \includegraphics[width=0.45\textwidth]{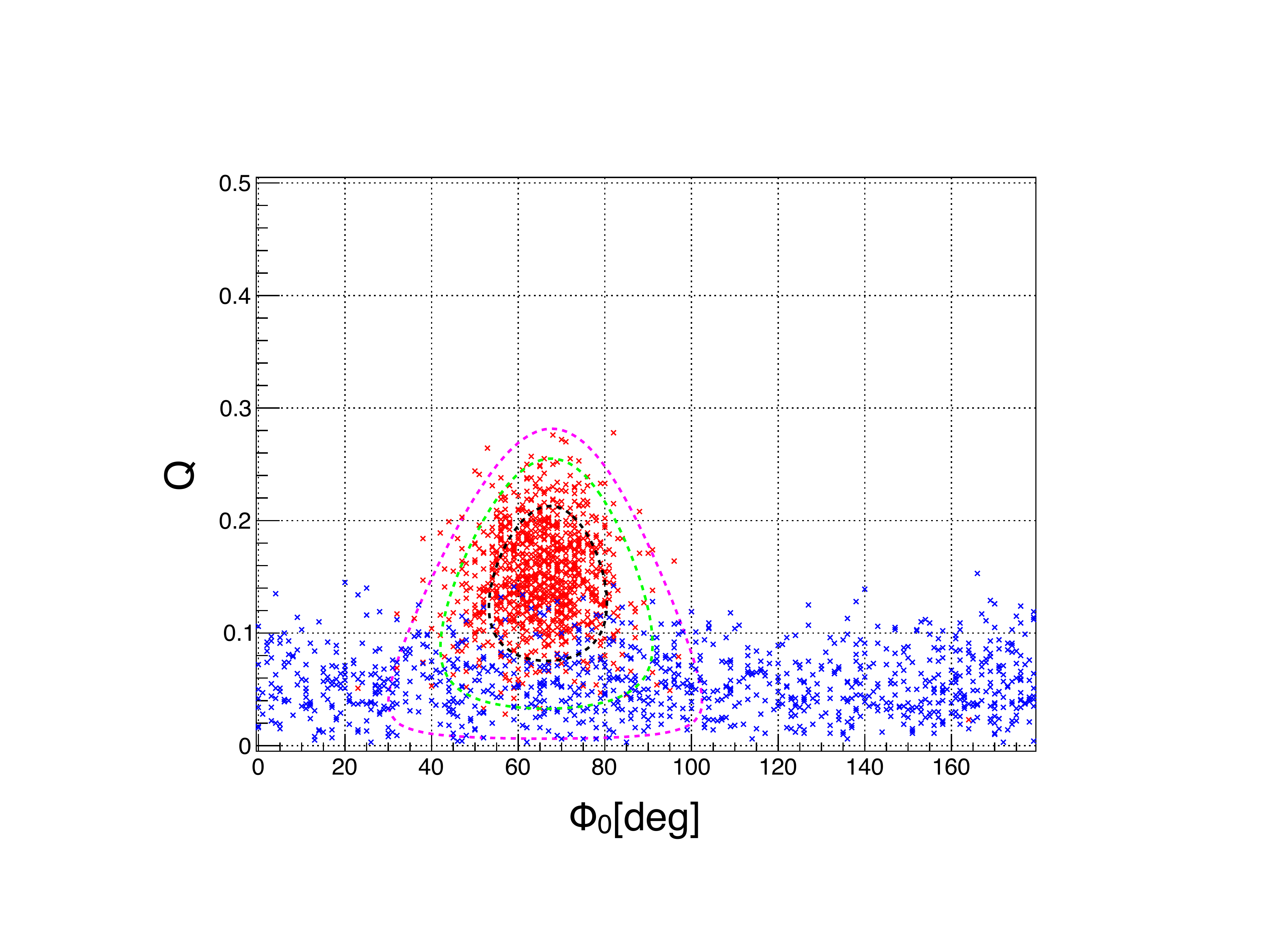} 
 \end{center}
\caption{%
The results of Likelihood estimations for 1000 sets of simulation data. 
The red points show the best-fit parameters for the Crab simulation data with the 
polarization parameters ($\Pi = 0.22$ and $\phi_0 = 67\degree$) derived from the observation data, and, the 
blue points show the best-fit parameters for the unpolarized simulation data. 
The contours are same as in Figure~\ref{fig:crab_obs_result}.
}\label{fig:obs_sim_1000times}
\end{figure}

The distribution of the best combinations of $Q$ and $\Phi_0$ from the fits for the 1000 sets of the Crab simulation data are shown as the 
red points of Figure~\ref{fig:obs_sim_1000times}.
The numbers of the data sets inside the contours of $\Delta \mathcal{L}$s 
of 2.30, 5.99, 9.21 are 668, 945 and 984, respectively. 
These numbers match the coverage probabilities in the case of two parameters.

\begin{figure}[ht]
 \begin{center}
  \includegraphics[width=0.4\textwidth]{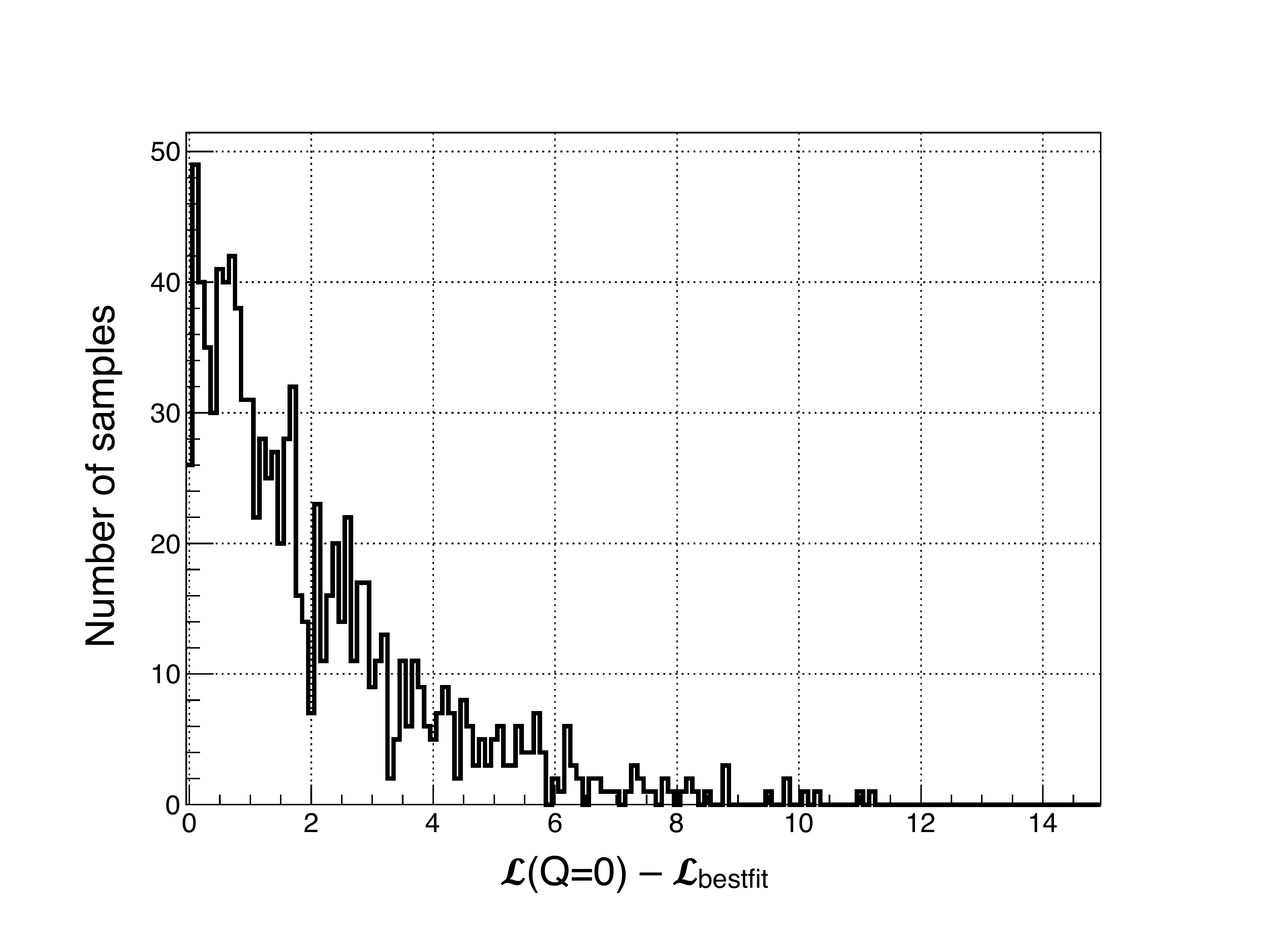} 
 \end{center}
\caption{The histogram of the difference between the minimum of the 
log likelihood $\mathcal{L}$ and the log likelihood of $Q = 0$ for the 1000 
sets of unpolarized simulation data.  The numbers of the data sets within the differences of 
2.30, 5.99 and 9.21 are 668, 955, and 993, respectively. 
The difference between the minimum of the log likelihood $\mathcal{L}$ and 
the log likelihood of $Q = 0$ also corresponds to the coverage probability for two parameters.
}
\label{fig:dist_deltaL_sim_1000times}
\end{figure}

In order to validate the confidence level for the detection of the polarized gamma-rays, 
we also prepared 1000 sets of unpolarized simulation data.  The results of the binned likelihood 
fits for the data sets are shown in the blue points of Figure~\ref{fig:obs_sim_1000times}.  
The distribution of the difference between the minimum of the log likelihood ($\mathcal{L}_0$) and 
the log likelihood of $Q = 0$ ($\mathcal{L}_{Q=0}$) is shown in Figure~\ref{fig:dist_deltaL_sim_1000times}.
It is confirmed that the value of the difference corresponds to the coverage probabilities 
in the case of two parameters.  Therefore, the $\Delta \mathcal{L}$ against the case of $Q = 0$ 
of 10.03 derived from the Crab observation corresponds to the confidence level of 99.3\%.

\begin{figure}[h]
 \begin{center}
  \includegraphics[width=0.4\textwidth]{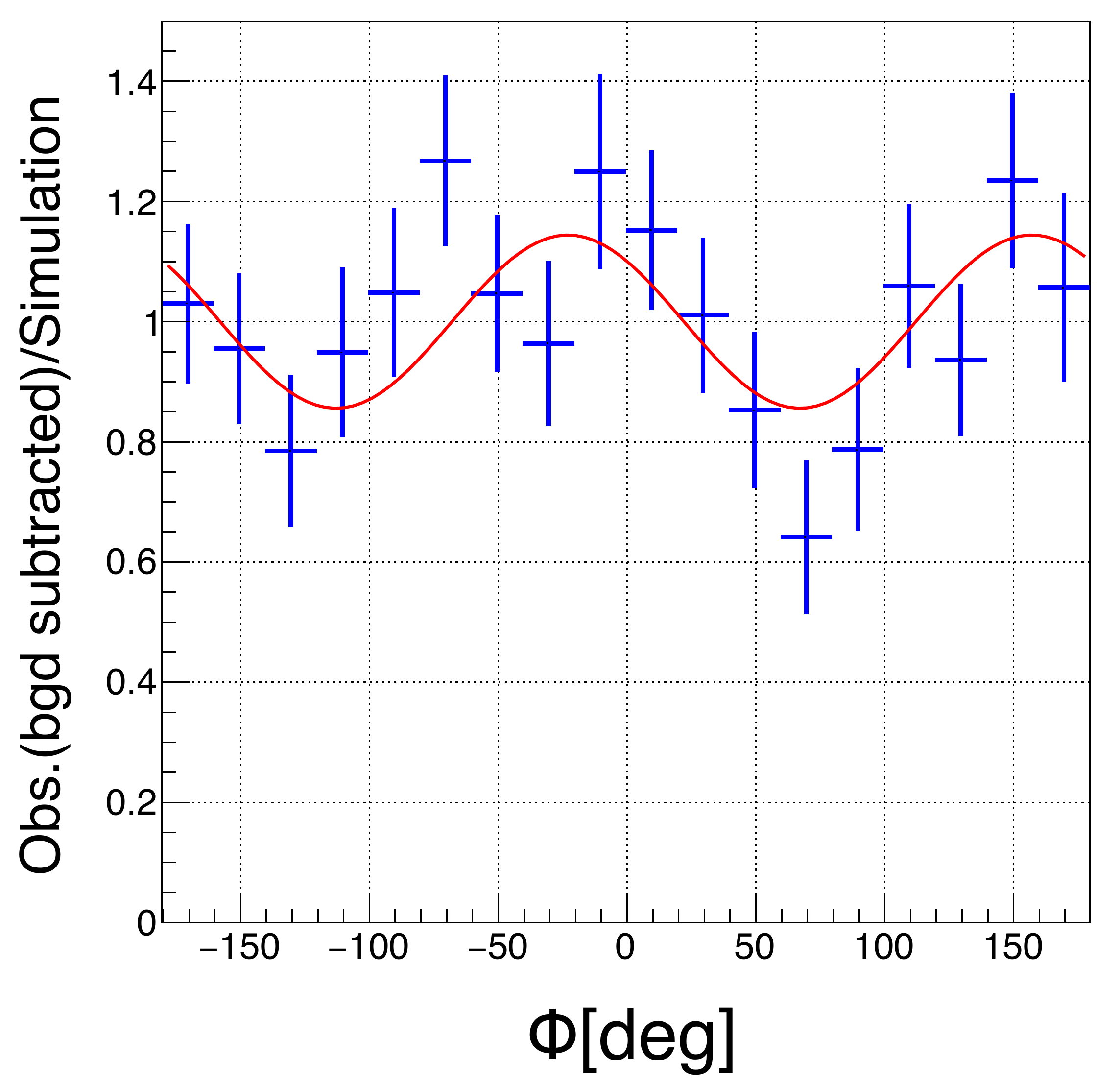} 
 \end{center}
\caption{Modulation curve of the Crab nebula observed with SGD. 
The data points show the ratio of the observation data subtracted the background to the unpolarized simulation data.
The error bar size indicates their statistical errors.
The red curve shows the sine curve function substituted the estimated parameters by the log-likelihood fitting.
}\label{fig:crab_mod}
\end{figure}

\begin{figure}[h]
 \begin{center}
  \includegraphics[width=0.4\textwidth]{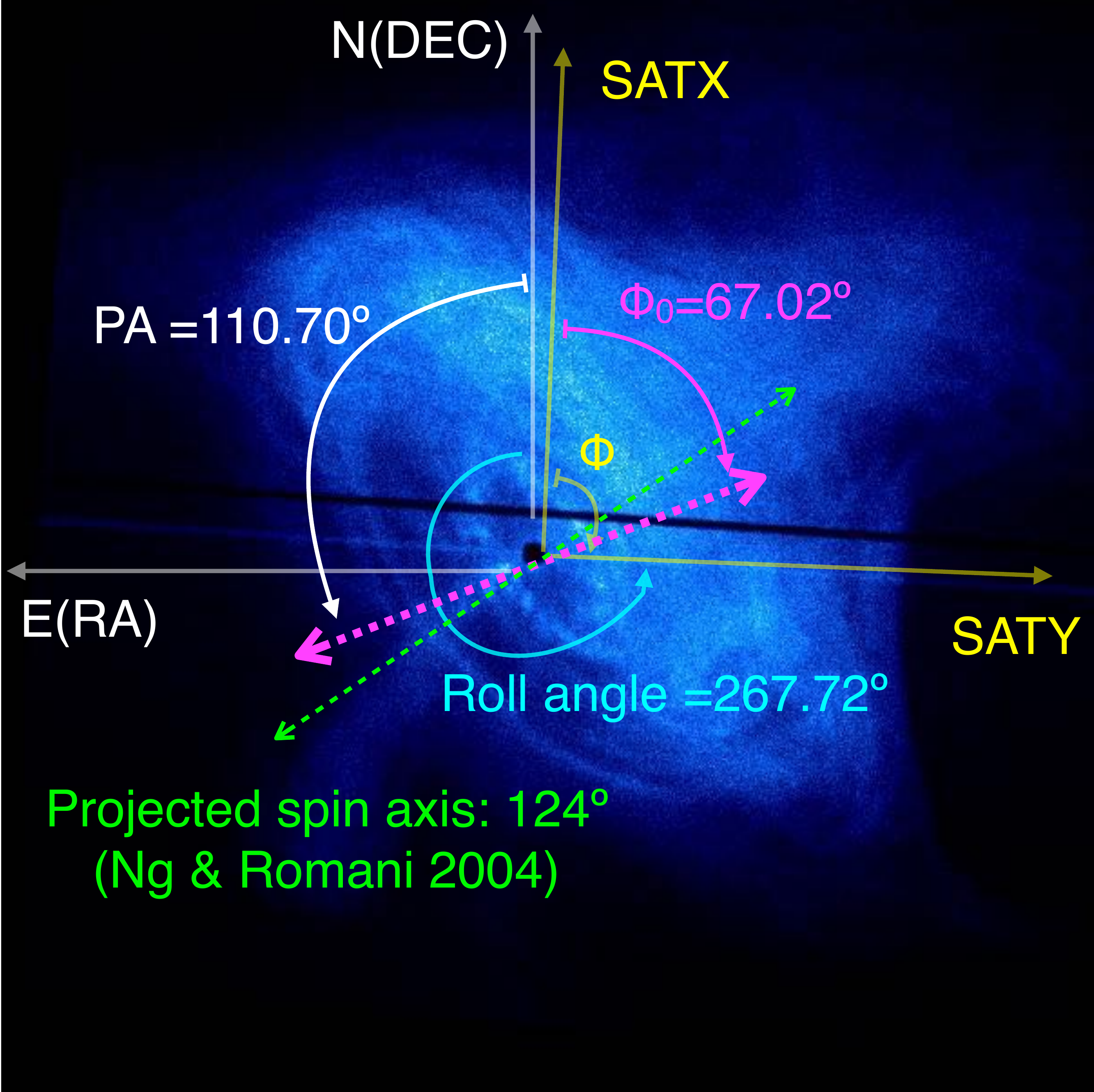} 
 \end{center}
\caption{The polarization angle of the gamma-rays from the Crab nebula determined by SGD. 
The direction of the polarization angle is drawn on the X-ray image of Crab with 
Chandra.}\label{fig:crab_image}
\end{figure}

Figure~\ref{fig:crab_mod} shows the phi distribution of the gamma rays from the Crab nebula with the 
parameters determined in this analysis.  
Figure~\ref{fig:crab_image} shows the relation between the satellite coordinate 
and the sky coordinate. The roll angle during the Crab observation was 267.72\degree~, and then, 
$\phi_0 = 67.02\degree$ corresponds the polarization angle of 110.70\degree~. 

\newpage

\section{Discussion}

\subsection{Comparison with other measurements}

The detection of polarization, and the measurement of 
its angle indicates the direction of an electric field vector of radiation.
In our analysis, the polarization angle is derived to be 
${\mathrm{PA}} = 110.7\degree\ \mbox{}^{+13.2\degree}_{-13.0\degree}$.  The energy range 
of gamma-rays contributing most significantly to this measurement is $\sim$~60--160~keV. 
All pulse phases of the Crab nebula emission are integrated.
The spin axis of the Crab pulsar is estimated $124.0\degree \pm 0.1\degree$ from X-ray imaging \citep{Ng:2004}.  
Therefore, the direction of the electric vector of radiation as measured by the SGD is about one standard deviation with the spin axis.  

The Crab polarization observation results from other instruments are listed in Table~\ref{tab:OtherResults}.
These instruments can be divided into three types based on the material of the scatterer.
The PoGO+ and the SGD employ carbon and silicon for as scatterer, respectively, while remaining instruments employ CZT or germanium.
Since the cross section of the Compton scattering exceeds that of the photo absorption at around 20~keV for carbon, around 60~keV for 
silicon and above 150~keV for germanium and CZT, which constrain the minimum energy range for each instrument.
Since the flux decreases with $E^{-2}$, the effective maximum energy for polarization measurements will be less than four times of the minimum energy.
Therefore, the PoGO+, the SGD and the other instruments have more or less non-overlapping energy range and are complimentary.
The PoGO$+$ team has reported the polarization angle ${\mathrm{PA}} = 131.3\degree \pm 6.8\degree$ and the 
polarization fraction ${\mathrm{PF}} = 20.9\%\pm5.0\%$
for the pulse-integrated, and ${\mathrm{PA}} = 137\degree \pm 15\degree$ and ${\mathrm{PF}} = 17.4\%^{+8.6\%}_{-9.3\%}$ for off-pulse period \citep{Chauvin:2017}.
Our results are consistent with the PoGO$+$ results.
On the other hand, for the higher energy range, the INTEGRAL IBIS, SPI and the AstroSat CZTI have performed 
the polarization observation of the Crab nebula in recent years, 
and, reported the slightly higher polarization fractions than our results.
Furthermore, the AstroSat CZTI reported varying polarization fraction during the off-peak period \citep{Vadawale:2017}.
However, we have not been able to 
verify those results because of extremely short observation time, which was less than 1/18th of the PoGO+, 
and less than 1/100th of the higher energy instrument.
Despite such short observation time, the errors of our measurements are within a factor of two of other instruments.
This result demonstrate the effectiveness of the SGD design such as high modulation factor of the azimuthal angle dependence, 
highly efficient instrument design and low backgrounds.
Extrapolating from this result, we expect that the 20~ks SGD observation can achieve statistical error equivalent 
with the PoGO+ and the AstroSAT CZTI, and the 80~ks SGD observation can perform phase resolved polarization measurements with similar errors.

\begin{table*}
  \tbl{Crab Polarization observation results}{%
  \begin{tabular}{ccccccc}
      \hline
Satellite/Instruments & Energy band       & Polarization angle [{}\degree] & Polarization fraction [\%] & Exposure time & phase & supplement \\ 
\hline
PoGO$+$(Balloon Exp.) & 20--160~keV       & $131.3\pm6.8$             & $20.9\pm5.0$             & 92 ks         & All   & \cite{Chauvin:2017}\\
Hitomi/SGD            & 60--160~keV       & $110.7^{+13.2}_{-13.0}$   & $22.1\pm10.6$            & 5 ks          & All   & this work \\
AstroSat/CZTI         & 100--380~keV      & $143.5\pm2.8$             & $32.7\pm5.8$             & 800 ks        & All   & \cite{Vadawale:2017} \\
INTEGRAL/SPI          & 130--440~keV      & $117\pm9$                 & $28\pm6$                 & 600 ks        & All   & \cite{Chauvin:2013}\\
INTEGRAL/IBIS         & 200--800~keV      & $110\pm11$                & $47^{+19}_{-13}$         & 1200 ks       & All   & \cite{Forot:2008} \\
\hline
    \end{tabular}}\label{tab:OtherResults}
\begin{tabnote}
\end{tabnote}
\end{table*}

\subsection{Implications on the source configuration}

We make a simple model of the polarization by assuming that the magnetic field is purely toroidal and the particle distribution function 
is isotropic (cf. \cite{Woltjer:1958}). The observed synchrotron radiation should be polarized along the projected symmetry axis (e.g. \cite{Rybicki:1979}). 
The degree of polarization depends upon the spectral index $\alpha\equiv-(1+d\ln N_x/d\ln E_X)$ where $N_X$ is the number of photons 
per unit photon energy $E_X$. It can be simply calculated by integration over azimuthal angle $\phi$ around a circular ring with axis inclined 
at an angle $\theta$ to the line of sight, $\bf n$. In these coordinates, $\hat{\bf B}\cdot\hat{\bf n}=\sin\theta\sin\phi$, 
and the angle $\chi$ between the local and and the mean polarization direction, satisfies $\cos2\chi=(\cos^2\phi-\cos^2\theta\sin^2\phi)/(1-\sin^2\theta\sin^2\phi)$. 
The mean degree of polarization is then given by
\begin{equation}
\Pi=\frac{(\alpha+1)\int_0^{2\pi}d\phi(1-\sin^2\theta\sin^2\phi)^{(\alpha-1)/2}(\cos^2\phi-\cos^2\theta\sin^2\phi)}{(\alpha+7/3)\int_0^{2\pi}d\phi(1-\sin^2\theta\sin^2\phi)^{(\alpha+1)/2}}.
\end{equation}
For the measured parameters, $\alpha=1.1$, $\theta=60\degree$, this evaluates to $\Pi=0.37$, The measured mean polarization is comfortably below this value suggesting that 
the magnetic field is moderately disordered relative to our simple model and the particle distribution function may be anisotropic. MHD and PIC simulations can be used to investigate this further.

\section{Conclusions}
The Soft Gamma-ray Detector (SGD) on board the Hitomi satellite observed the Crab nebula during the initial test observation period of Hitomi.
Even though this observation was not intended for the scientific analyses, the gamma-ray radiation from the Crab nebula was detected by combining the careful data analysis, 
the background estimation, and the SGD Monte Carlo simulations.
Moreover, polarization measurements were performed for the data obtained with SGD Compton cameras, 
and, polarization of soft gamma-ray emission was successfully detected.
The obtained polarization fraction of the phase-integrated Crab emission (sum of pulsar and nebula emissions) 
was 22.1\% $\pm$ 10.6\% and, the polarization angle was 110.7\degree +13.2\degree/$-$13.0\degree~ 
(The errors correspond to the 1 sigma deviation) despite extremely short observation time of 5~ks.
The confidence level of the polarization detection was 99.3\%.
This is well-described as the soft gamma-ray emission arising predominantly from energetic particles radiating 
via the synchrotron process in the toroidal magnetic field in the Crab nebula, 
roughly symmetric around the rotation axis of the Crab pulsar.
This result demonstrates that the SGD design is highly optimized for polarization measurements.

\newpage

\begin{ack}
We thank all the JAXA members who have contributed to the ASTRO-H (Hitomi) project. 
All U.S. members gratefully acknowledge support through the NASA Science Mission Directorate. 
Stanford and SLAC members acknowledge support via DoE contract to SLAC National Accelerator Laboratory DE-AC3-76SF00515 and NASA grant NNX15AM19G. 
Part of this work was performed under the auspices of the US DoE by LLNL under Contract DE-AC52-07NA27344. 
Support from the European Space Agency is gratefully acknowledged. 
French members acknowledge support from CNES, the Centre National d'\'{E}tudes Spatiales. 
SRON is supported by NWO, the Netherlands Organization for Scientific Research. 
The Swiss team acknowledges the support of the Swiss Secretariat for Education, Research and Innovation (SERI). 
The Canadian Space Agency is acknowledged for the support of the Canadian members. 
We acknowledge support from JSPS/MEXT KAKENHI grant numbers 15H00773, 15H00785, 15H02090, 15H03639, 15H05438, 15K05107, 15K17610, 15K17657, 16J02333, 16H00949, 16H06342, 16K05295, 16K05296, 16K05300, 16K13787, 16K17672, 16K17673, 17H02864, 17K05393, 17J04197, 21659292, 23340055, 23340071, 23540280, 24105007, 24244014. 24540232, 25105516, 25109004, 25247028, 25287042, 25287059, 25400236, 25800119, 26109506, 26220703, 26400228, 26610047, 26800102, 26800160, JP15H02070, JP15H03641, JP15H03642, JP15H06896, JP16H03983, JP16K05296, JP16K05309, and JP16K17667. 
The following NASA grants are acknowledged: NNX15AC76G, NNX15AE16G, NNX15AK71G, NNX15AU54G, NNX15AW94G, and NNG15PP48P to Eureka Scientific. 
This work was partly supported by Leading Initiative for Excellent Young Researchers, MEXT, Japan, and also by the Research Fellowship of JSPS for Young Scientists. 
H. Akamatsu acknowledges the support of NWO via a Veni grant. 
C. Done acknowledges STFC funding under grant ST/L00075X/1. 
A. Fabian and C. Pinto acknowledge ERC Advanced Grant 340442. 
P. Gandhi acknowledges a JAXA International Top Young Fellowship and UK Science and Technology Funding Council (STFC) grant ST/J003697/2. 
Y. Ichinohe, K. Nobukawa, and H. Seta are supported by the Research Fellow of JSPS for Young Scientists program. 
N. Kawai is supported by the Grant-in-Aid for Scientific Research on Innovative Areas “New Developments in Astrophysics Through Multi-Messenger Observations of Gravitational Wave Sources.” 
S. Kitamoto is partially supported by the MEXT Supported Program for the Strategic Research Foundation at Private Universities, 2014--2018. 
B. McNamara and S. Safi-Harb acknowledge support from NSERC. T. Dotani, T. Takahashi, T. Tamagawa, M. Tsujimoto, and Y. Uchiyama acknowledge support from the Grant-in-Aid for Scientific Research on Innovative Areas “Nuclear Matter in Neutron Stars Investigated by Experiments and Astronomical Observations.” 
N. Werner is supported by the Lend\"ulet LP2016-11 grant from the Hungarian Academy of Sciences. 
D. Wilkins is supported by NASA through Einstein Fellowship grant number PF6-170160, awarded by the Chandra X-ray Center and operated by the Smithsonian Astrophysical Observatory for NASA under contract NAS8-03060.

We thank contributions by many companies, including in particular NEC, Mitsubishi Heavy Industries, Sumitomo Heavy Industries, Japan Aviation Electronics Industry, 
Hamamatsu Photonics, Acrorad, Ideas, SUPER RESIN and OKEN. 
We acknowledge strong support from the following engineers. JAXA/ISAS: Chris Baluta, Nobutaka Bando, Atsushi Harayama, Kazuyuki Hirose, Kosei Ishimura, Naoko Iwata, Taro Kawano, Shigeo Kawasaki, Kenji Minesugi, Chikara Natsukari, Hiroyuki Ogawa, Mina Ogawa, Masayuki Ohta, Tsuyoshi Okazaki, Shin-ichiro Sakai, Yasuko Shibano, Maki Shida, Takanobu Shimada, Atsushi Wada, and Takahiro Yamada; JAXA/TKSC: Atsushi Okamoto, Yoichi Sato, Keisuke Shinozaki, and Hiroyuki Sugita; Chubu University: Yoshiharu Namba; Ehime University: Keiji Ogi; Kochi University of Technology: Tatsuro Kosaka; Miyazaki University: Yusuke Nishioka; Nagoya University: Housei Nagano; NASA/GSFC: Thomas Bialas, Kevin Boyce, Edgar Canavan, Michael DiPirro, Mark Kimball, Candace Masters, Daniel Mcguinness, Joseph Miko, Theodore Muench, James Pontius, Peter Shirron, Cynthia Simmons, Gary Sneiderman, and Tomomi Watanabe; ADNET Systems: Michael Witthoeft, Kristin Rutkowski, Robert S. Hill, and Joseph Eggen; Wyle Information Systems: Andrew Sargent and Michael Dutka; Noqsi Aerospace Ltd: John Doty; Stanford University/KIPAC: Makoto Asai and Kirk Gilmore; ESA (Netherlands): Chris Jewell; SRON: Daniel Haas, Martin Frericks, Philippe Laubert, and Paul Lowes; University of Geneva: Philipp Azzarello; CSA: Alex Koujelev and Franco Moroso.
Finally, we greatly appreciate the dedicated work by all students in participating institutions.
\end{ack}

\section*{Author contributions}
S. Watanabe led this study in data analysis and writing drafts. 
He also contributed to the SGD overall design, fabrication, integration and tests, in-orbit operations and calibration.
Y. Uchida contributed to data analysis and preparing the drafts in addition to the SGD calibration.
H. Odaka worked for the SGD Monte Carlo simulator and contributed to data analysis.
G. Madejski contributed to writing drafts and improved the draft. 
K. Hayashi contributed to the testing and calibration of the SGD and the SGD operations.
T. Mizuno contributed to data analysis in addition to the polarized beam experiment.
R. Sato and Y. Yatsu worked for the SGD BGO shield design, fabrication and testing, and also contributed the Hitomi's operations.

\appendix 
%
%
%


\end{document}